\documentclass[final,authoryear,5p,times,twocolumn]{elsarticle}

\usepackage[utf8]{inputenc}
\usepackage{graphicx}
\usepackage{amssymb}
\usepackage{amsmath}
\usepackage{journals}
\usepackage[justification=justified,singlelinecheck=false]{caption}
\usepackage[usenames,dvipsnames]{xcolor}
\usepackage[colorlinks=True,
	    urlcolor=blue,
	    linkcolor=blue,
	    citecolor= blue,
	    pdftex]{hyperref}
\usepackage{cleveref}
\usepackage{longtable}

\newcommand{\tablecases}[1]{#1}

\newcommand{\noc}[1]{#1}

\begin{document}

%\begin{linenumbers}

\begin{frontmatter}

\title{Anelastic dynamo models with variable electrical conductivity: an application to gas giants}
\author{Lúcia D. V. Duarte\fnref{label2,label3}}
\ead{duarte@mps.mpg.de}
\author{Thomas Gastine\fnref{label2}}
\author{Johannes Wicht\fnref{label2}}

\fntext[label2]{Max-Planck-Institut für Sonnensystemforschung}
\fntext[label3]{Technische Universität Braunschweig}
\address{Max-Planck-Str. 2, 37191 Katlenburg-Lindau, Germany}

\begin{abstract}

The observed surface dynamics of Jupiter and Saturn \noc{are} dominated by a banded
system of fierce zonal winds. The depth of these winds remains unclear but they are
thought to be confined to the very outer envelopes where hydrogen
remains molecular and the electrical conductivity is small.
The dynamo maintaining the dipole-dominated magnetic fields
of both gas giants, on the other hand, likely operates in the deeper
interior where hydrogen assumes a metallic state.

Here, we present numerical simulations that attempt to model both the
zonal winds and the interior dynamo action in an integrated approach.
Using the anelastic version of the MHD code MagIC, we explore the effects
of density stratification and radial electrical conductivity variations.
The electrical conductivity is mostly assumed to remain constant in the thicker
inner metallic region and it decays exponentially towards the outer
boundary throughout the molecular envelope.

Our results show that the combination of a stronger density stratification and
a weaker conducting outer layer is essential for reconciling dipole dominated
dynamo action and a fierce equatorial zonal jet. Previous simulations with
homogeneous electrical conductivity show that both are \noc{mutually} exclusive,
with solutions either having strong zonal winds and multipolar magnetic
fields or weak zonal winds and dipole-dominated magnetic fields.
All jets tend to be geostrophic and therefore reach right through the
convective shell in our simulations.

The particular setup explored here allows a strong equatorial jet to
remain confined to the weaker conducting outer region where it does not interfere
with the deeper seated dynamo action.
The flanking mid to high latitude jets, on the other hand, have to
remain faint to yield a strongly dipolar magnetic field.
The fiercer jets on Jupiter and Saturn only seem compatible
with the observed dipolar fields when they remain confined to a
weaker conducting outer layer.

\end{abstract}


\end{frontmatter}

\section{Introduction}
\label{intro}

The gas giants, Jupiter and Saturn, mainly consist
of a hydrogen-helium mixture.
Due to the large pressures and temperatures reached inside these planets,
hydrogen acquires metallic properties \citep{Chabrier92,Fortney10}.
The transition happens at 85--90\% of
Jupiter's and $65$\% of Saturn's radii.
A classical view is that the lower metallic layer likely hosts
the dynamo of these planets, while the upper molecular envelope
accommodates the observed fierce zonal jets. Higher densities,
Lorentz forces and Ohmic diffusion would lead to a more sluggish dynamics
in the metallic layer and confine the zonal winds to the upper region.
Traditional dynamical models therefore treat the two layers separately
with dynamo simulations modelling only the metallic layer and jet
simulations concentrating on the molecular envelope.

The zonal jets have been investigated since the 70s by tracking
cloud features
(see, for example, \citeauthor{Ingersoll79}, \citeyear{Ingersoll79} for Jupiter
and \citeauthor{SanchezLavega82}, \citeyear{SanchezLavega82}
for Saturn). Their driving forces and depth are still debated. Some authors
argue that they are a shallow weather phenomenon \citep{Williams78,Cho96}
while others promote deeper-rooted jets that extend through the whole
molecular envelope \citep{Heimpel05,Jones09,Gastine12}.
Both gas giants emit roughly twice as much energy as they receive from
the sun which implies vigorous interior convection. In the rotationally-dominated
dynamics ruling planetary atmospheres, interior convection %quite
naturally drives zonal winds via Reynolds stresses \citep[i.e. a statistical correlation
between the convective flow components;][]{Christensen02,Heimpel05}.
These winds follow a geostrophic structure, minimizing variations in the direction
of the rotation axis, and therefore reach through the whole fluid
atmosphere.
\cite{Lian08} show that even when the forcing is restricted to a shallow
weather layer the jets may reach much deeper into the planet.
\cite{Kaspi09}, on the other hand, present an anelastic deep convection model
where the equatorial \noc{zonal flow} is geostrophic and the higher latitude jets are confined to
the outer few percent in radius.

Saturn's magnetic field is very axisymmetric and strongly
concentrated at higher latitudes \citep{Cao12} which is
incompatible with the results of a classical Earth-like dynamo model.
A stably stratified layer at the top of the dynamo region
\citep{Christensen08,Stanley10} or a completely different dynamo driven
by differential rotation \citep{Cao12} are two proposed alternatives for
the special situation encountered at Saturn.

Here we concentrate on Jupiter whose field is very similar to the
geomagnetic field so that the well-explored geodynamo models also
seem to apply at first sight.
These models typically adopt the Boussinesq approximation where the
mild $30$\% density stratification of Earth's core is simply ignored.
In Jupiter, however, the density increases by more than
a factor of 5000 below the $1\,$bar level. While the stratification is mostly
concentrated in the outer molecular envelope, the density still rises by about
one order of magnitude across the metallic layer \citep[Fig.~1
of][]{French12}.
Some newer numerical models therefore use the anelastic approximation which
allows to incorporate the effects of the background density stratification
while filtering out fast sound waves \citep{Glatz123,Stanley09,Jones09}.

In an extensive parameter study, \cite{Gastine12a} (hereafter referred to as GDW12)
show that dipole-dominated
dynamos are rather rare when stronger stratifications are assumed.
GDW12 quantify the stratification in their anelastic models in terms of
the number of density scale heights $N_\rho=\ln(\rho_i/\rho_o)$,
where $\rho_i$ and $\rho_o$ are
the densities at the inner and outer boundaries of the simulated shell,
respectively.
For the larger density stratifications $N_\rho\!>\!2$, a value that corresponds to
an increase by a factor $7.4$, no dipole-dominated solutions were found.
This is attributed to the fact that the focus of convective action moves
progressively outward in cylindrical radius when the stratification is
intensified.
Once the convective columns are mainly confined to a relatively thin
outer shell, a non-axisymmetric dynamo mode is preferred
that has previously only been observed in mean field dynamo
simulations \citep{Rudiger03,Jiang06}. We will refer to this as the
thin-shell dynamo model in the following.

For the smaller to intermediate stratifications $N_\rho\!\le\!2$, GDW12 find
dipole dominated magnetic fields when the local Rossby number remains smaller
than a critical value of $Ro_{\ell c}\!\approx\!0.1$. This is consistent with the
findings of \cite{Christensen06} who introduced $Ro_\ell$ as
a measure for the relative importance of inertia
in their Boussinesq models (see Eq. \ref{eq:rol}).
Multipolar solutions with weaker magnetic fields on the other hand exist for all
$Ro_\ell$ values which means that both types of solutions coexist
below $Ro_{\ell c}$ for identical model parameters, forming two distinct branches.
This so-called bistability can be attributed to the fact that free-slip
boundary conditions were employed \citep{Simitev09,Schrinner12,Gastine12a}.
These conditions allow strong zonal winds to develop that
compete with large scale magnetic fields. On the dipolar branch, zonal winds
are weak, on the multipolar branch they are stronger.
When no-slip conditions are used zonal flows generally remain weaker
and only the dipolar branch is found for
$Ro_\ell<Ro_{\ell c}$ \citep{Christensen06}.

\textit{Ab initio} calculations suggest that there is actually no clear
phase transition between the regions of molecular and metallic hydrogen states
\citep{Lorenzen11,French12}. In the dynamo context, the electrical
conductivity profile is of particular importance.
Due to the increasing degree of hydrogen ionization, the conductivity
rises super-exponentially with depth and matches the conductivity of
the metallic region at the transition radius without any pronounced jump.
The classical separation of the dynamics for the two envelopes thus becomes
questionable. \cite{Liu08} argue that this has important consequences
for the depth of the zonal winds which should remain confined
to a shallow outer layer where the conductivity remains negligible.
The strong shear associated with the zonal winds would otherwise create strong
azimuthal magnetic field and lead to Ohmic heating incompatible
with the observed luminosity \citep[see however][]{Glatz08}.

\cite{Stanley09} present an anelastic simulation of a relatively
thin shell with exponentially decaying electrical conductivity to model
the very outer part of the shell.
The model uses extreme parameters
(i.e. low Ekman and Prandtl number and high Rayleigh number) and a
dipole-dominated magnetic field develops in the presence of strong
geostrophic zonal winds.
However, since a detailed discussion and a systematic parameter study are missing,
it remains impossible to disentangle the effects of density
stratification, varying conductivity, and the particular parameter choice.
\cite{Gomez10} and \cite{Heimpel11} also include a radial conductivity profile
in their deep shell Boussinesq models, with a constant conductivity
in the deeper interior and an exponential decay in the outer part.
These models also demonstrate that well-pronounced deep-rooted zonal winds
can be compatible with dipole-dominated dynamo action.

The present paper extends the work of GDW12 by adding
an electrical conductivity profile
loosely based on the \textit{ab initio} calculations
by \cite{French12}. Following \cite{Gomez10} and \cite{Heimpel11}, the electrical
conductivity profile assumes a constant value in the metallic region and an
exponential decay in the molecular region.
The aim is to systematically explore under which
circumstances dipole-dominated dynamo
action and strong zonal surface winds can coexist in anelastic dynamo models.

We describe our model in section \ref{model}
with special attention to the anelastic
formulation and the electrical conductivity profile.
The numerical results are presented in section \ref{results}, first
concentrating on the question of dipole-dominance and then on the
dynamo mechanism. Section \ref{conclusions} summarizes
our main results and discusses their implications for the gas giants.

\section{Model}
\label{model}

\subsection{Anelastic approximation}
\label{anelasticeqs}

The fluid and convective interior of the planet is modelled by solving the MHD
equations in a rapidly-rotating spherical shell. Previous models typically used the
Boussinesq approximation, which neglects the background
density and temperature variations. This is questionable in gas planets and,
following \cite{Glatz123}, \cite{Brag95} and \cite{Lantz99}, we therefore adopt
the anelastic approximation. This allows to include background variations
while ruling out sound waves by neglecting fast local density variations.

We solve the equations in a dimensionless form \citep[e.g.][]{Christensen06},
using the shell thickness $d=r_o-r_i$ as a length scale and
the viscous diffusion time $\tau_{\nu}=d^2/{\nu}$ as a timescale.
Here, $r_o$ and $r_i$ are the outer and inner radii, respectively, and
$\nu$ is the kinematic viscosity.
Temperature and density are both
non-dimensionalized by their values at the outer boundary, $T_o$ and $\rho_o$.
We employ constant entropy boundary conditions and use the imposed
contrast $\Delta s$ across the shell as the entropy scale. There are no internal heat sources and
all the heating coming into the shell via the inner boundary leaves it through the outer.
While this is not the most realistic heating mode for gas giants, it has been chosen to
ease the comparisons with more classical Boussinesq simulations.
The magnetic field is scaled by $\sqrt{\Omega\mu\lambda_i\rho_o}$,
where $\Omega$ is the rotation rate of the shell and
$\lambda_i$ is the inner boundary reference value of the magnetic
diffusivity $\lambda(r)=1/(\sigma(r)\mu)$.
Here, $\mu$ is the magnetic permeability and $\sigma(r)$ is
the prescribed electrical conductivity profile.
Below we will also use the normalized magnetic diffusivity
and electrical conductivity profiles related via:
$\tilde{\lambda}(r)=\lambda(r)/\lambda_i=\sigma_i/\sigma(r) =
\tilde{\sigma}(r)^{-1}$.

The medium is assumed to be an electrically conducting ideal polytropic gas.
Generally, dynamo simulations solve for small variations around
an adiabatic hydrostatic background state that we mark with a
tilde in the following. The background temperature profile is then
defined by the background temperature gradient
$d\tilde{T}/dr=-g(r)/c_p$ and the density profile by $\tilde{\rho}(r)=
\tilde{T}^m$, where $m$ is the polytropic index.
For simplicity,
we adopt a gravity profile proportional to radius which implicitly assumes a
homogeneous density. The other extreme is to assume that all the mass is concentrated
in the centre, which leads to a gravity profile %$\propto 1/r^2$ \citep{Jones11}. \cite{Gastine12}
proportional to $1/r^2$ \citep{Glatz123,Jones11}. GDW12 show that both gravity profiles
lead to very similar results. The true profile of the gas giants lies somewhere in-between.
The temperature reference state is then given by

\begin{equation}
\begin{split}
      \tilde{T}(r)=-c_0\bigg(\frac{r}{r_o}\bigg)^2+1+c_0 \textrm{,}
\end{split}
\label{eq:adiabatrhoeq}
\end{equation}
where
\begin{equation}
      c_0=\frac{(e^{\frac{N_{\rho}}{m}}-1)}{(1-{\eta}^2)} \textrm{.}
\label{eq:constants}
\end{equation}

$N_{\rho}=\ln(\rho_i /\rho_o)$ is the number of density scale heights
between the inner and the outer boundaries of the shell and $\eta$ is the ratio
between the corresponding radii
\citep[see][for the full derivation of the reference state]{Jones11,Gastine12}.

The dimensionless form of the anelastic equations is

\begin{equation}
\begin{split}
      E\,\bigg(\frac{\partial \mathbf{u}}{\partial t} + \mathbf{u}\cdot\nabla \mathbf{u}\bigg)
      = - \nabla\frac{p}{\tilde{\rho}} - 2\mathbf{e}_z\times\mathbf{u}
      + \frac{Ra\,E}{Pr}\frac{r}{r_o}s\,\mathbf{e}_r \\
      + \frac{1}{Pm_i\,\tilde{\rho}}(\nabla\times\mathbf{B})\times\mathbf{B}
      + \frac{E}{\tilde{\rho}}\nabla\cdot\textsf{S} \textrm{,}
\end{split}
\label{eq:navierstokeseq}
\end{equation}

\begin{equation}
      \frac{\partial \mathbf{B}}{\partial t} = \nabla\times(\mathbf{u}\times\mathbf{B})
      - \frac{1}{Pm_i}\nabla\times(\tilde{\lambda}\nabla\times\mathbf{B}) \textrm{,}
\label{eq:inductioneq}
\end{equation}

\begin{equation}
\begin{split}
      \tilde{\rho}\,\tilde{T}\,\bigg(\frac{\partial s}{\partial t} + \mathbf{u}\cdot\nabla s\bigg)
      = \frac{1}{Pr}\nabla\cdot (\tilde{\rho}\tilde{T}\nabla s) \\
      + \frac{Pr}{Ra}(1-\eta)c_0 Q_{\nu}
      + \frac{Pr}{Pm_i^2\,Ra\,E}(1-\eta)c_0 Q_j \textrm{,}
\end{split}
\label{eq:energyeq}
\end{equation}

\begin{equation}
      \nabla\cdot (\tilde{\rho}\mathbf{u}) = 0 \textrm{,}
\label{eq:divfloweq}
\end{equation}

\begin{equation}
      \nabla\cdot\mathbf{B} = 0 \textrm{.}
\label{eq:divfieldeq}
\end{equation}

The traceless rate-of-strain tensor $\textsf{S}$ for the homogeneous kinematic viscosity
assumed here is given by
\begin{equation}
      \textsf{S} = 2\tilde{\rho}\bigg[\textsf{e}_{ij}-\frac{1}{3}\delta_{ij}\nabla\cdot\mathbf{u}\bigg] \;\;\;\textrm{  and   }\;\;\;
      \textsf{e}_{ij} = \frac{1}{2}\bigg(\frac{\partial u_i}{\partial x_j}+\frac{\partial u_j}{\partial x_i}\bigg) \textrm{,}
\label{eq:straintensor}
\end{equation}
where $\delta_{ij}$ is the identity matrix.
The viscous and ohmic heating contributions are
\begin{equation}
      Q_{\nu}=2\tilde{\rho}\bigg[\textsf{e}_{ij}\textsf{e}_{ji}-\frac{1}{3}(\nabla\cdot\mathbf{u})^2\bigg] %\textrm{,}
\label{eq:vischeat}
\end{equation}
and
\begin{equation}
       Q_j=\tilde{\lambda}(\nabla\times\mathbf{B})^2 \textrm{.}
\label{eq:ohmicheat}
\end{equation}

The system of Eqs.~(\ref{eq:navierstokeseq}--\ref{eq:divfieldeq}) is governed by the
dimensionless Ekman number $E$, Rayleigh number $Ra$, Prandtl number $Pr$ and
magnetic Prandtl number at the inner boundary $Pm_i$:

\begin{equation}
      E=\frac{\nu}{\Omega d^2} \textrm{,}
\label{eq:ekmunber}
\end{equation}

\begin{equation}
      Ra=\frac{g_o d^3\Delta s}{c_p\nu\kappa} \textrm{,}
\label{eq:ramunber}
\end{equation}

\begin{equation}
      Pr=\frac{\nu}{\kappa} \textrm{,}
\label{eq:prmunber}
\end{equation}

\begin{equation}
      Pm_i=\frac{\nu}{\lambda_i} \textrm{.}
\label{eq:pmmunber}
\end{equation}
The specific heat $c_p$, the thermal diffusivity $\kappa$, magnetic
diffusivity $\lambda$ and kinematic viscosity $\nu$ are all assumed
to be homogeneous.
To quantify gravity we use the reference value $g_o$ at the outer boundary.

\subsection{Variable conductivity}
\label{varcond}

To simulate the variable electrical conductivity of hydrogen in the interior
of Jupiter, we employ a profile that corresponds to a constant
conductivity in the metallic hydrogen layer and an exponential decay in the outer
molecular envelope. Both branches are matched via a polynomial that also
ensures that the first radial derivative is continuous:
\begin{equation}
      \tilde{\sigma}(r) =
      \left\{
      \begin{array}{ll}
            1+(\tilde{\sigma}_m-1)\,\Bigg(\dfrac{r-r_i}{r_m-r_i}\Bigg)^a & r<r_m \\
            \tilde{\sigma}_m
            \exp{}\left[a\,\frac{r-r_m}{r_m-r_i}\,
            \frac{\tilde{\sigma}_m-1}{\tilde{\sigma}_m}\right] & r\ge r_m
      \end{array} \right.
       \textrm{.}
\label{varcondeq}
\end{equation}
The exponential decay with a rate $a$ starts at a radius $r_m$ where
the normalized conductivity has already decreased from $\tilde{\sigma}_i\!=\!1$
to $\tilde{\sigma}_m$. For convenience we also define the
relative transition radius in percentage: $\chi_m\!=\!r_m/r_o$.

\begin{figure}[h!]
\begin{center}
{\centering
      \includegraphics[trim=0cm 1cm 0.5cm 0cm, height=2.4in]{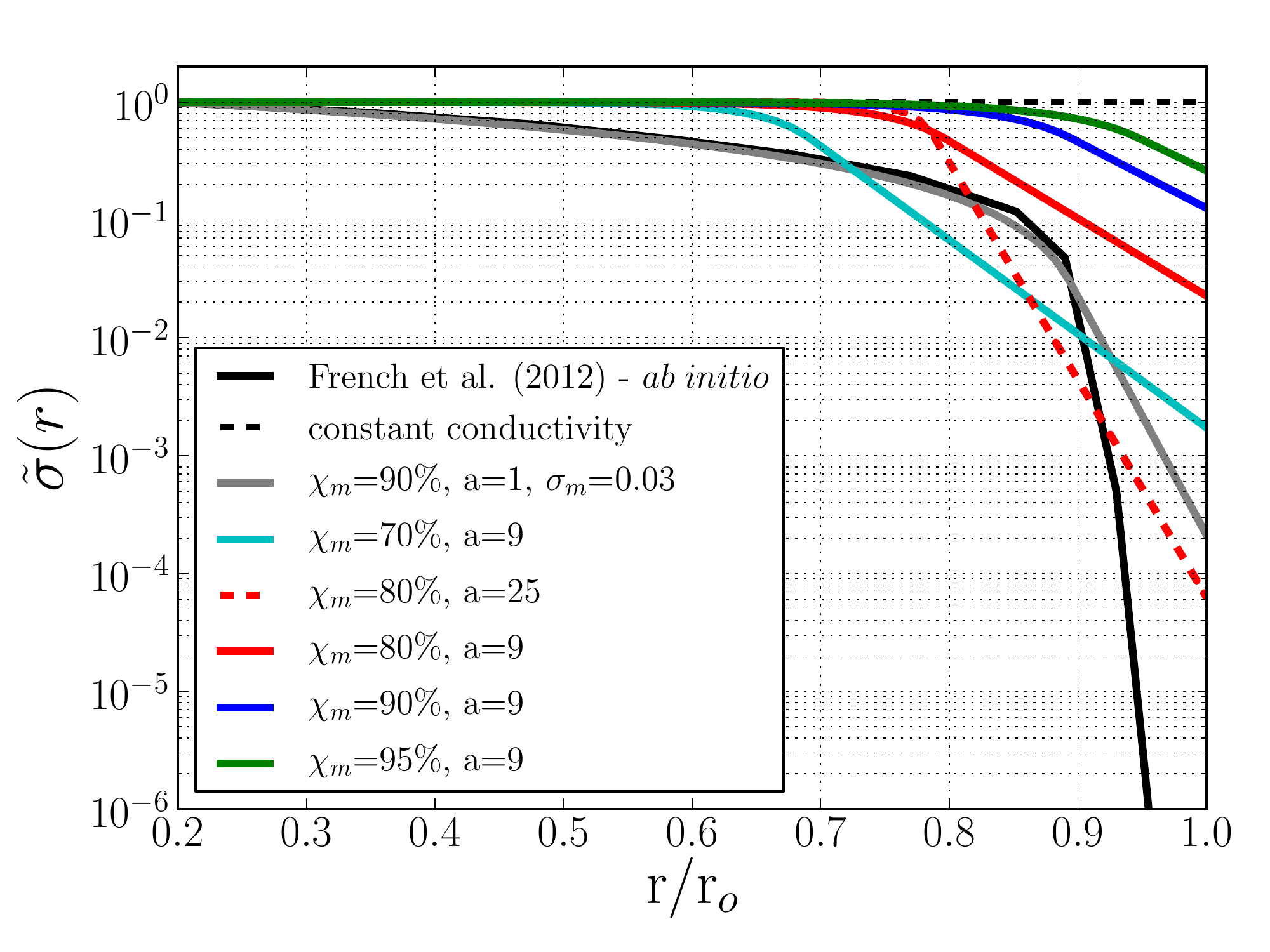}
      }
\caption{\small Radial profiles of electrical conductivity, used in this work. The black line
corresponds to the \textit{ab initio} solution from \cite{French12}. All the profiles
in colour, with either $a\!=\!9$ or $a\!=\!25$, have $\sigma_m\!=\!0.5$.
\label{fig:vconds}}
\end{center}
\end{figure}

This profile has first been used by \cite{Gomez10}
and it seems a fair first approximation to the results from
\textit{ab initio} calculations by \cite{French12}.
The super-exponential increase of electrical conductivity over the
molecular layer is \noc{not feasible} to model numerically (see Fig.~\ref{fig:vconds}).
We thus mainly use a rate of $a=9$ for our simulations, but we also tested
$a\!=\!25$ in a few cases (see Tab.~\ref{Tab2}) and $a\!=\!1$ for one case
with a different $\sigma_m$ (grey profile in Fig.~\ref{fig:vconds}, discussed in section
\ref{diporflow}).
In all the other cases, $\tilde{\sigma}_m$ was fixed to $0.5$ and
$\chi_m$ was varied assuming values of $95$, $90$, $80$ and $70$\%.
Corresponding simulations for homogeneous conductivity with $\chi_m\!=\!100\%$ can
be found in GDW12.

\subsection{Numerical model}
\label{anelasticmodel}

For the numerical simulation of the model described above, we use the anelastic
version of the MagIC code \citep{Wicht02,Gastine12}. This is a pseudo-spectral code
that solves Eqs.~(\ref{eq:navierstokeseq}--\ref{eq:divfieldeq}) in a spherical
shell using a poloidal/toroidal decomposition of the vector fields
$\tilde{\rho}\mathbf{u}$ and $\mathbf{B}$:

\begin{equation}
\begin{split}
      \tilde{\rho}\mathbf{u}=(\tilde{\rho}\mathbf{u})_{pol}+(\tilde{\rho}\mathbf{u})_{tor}
      =\nabla\times (\nabla\times w\,\mathbf{e}_r)+\nabla\times z\,\mathbf{e}_r \\
      \mathbf{B}=\mathbf{B}_{pol}+\mathbf{B}_{tor}=\nabla\times (\nabla\times c\,\mathbf{e}_r)+
      \nabla\times a\,\mathbf{e}_r \textrm{.}
\end{split}
\label{poloidaltoroidal}
\end{equation}
For the spectral representation of the dependence on latitude $\theta$
and longitude $\phi$,
the poloidal potentials $w$ and $c$, the toroidal potentials $z$ and
$a$, the entropy $S$ and the pressure $p$ are expanded in spherical harmonic
functions up to degree and order $\ell_{max}$.
Chebyshev polynomials up to degree $N_r$ are used in the radial direction.

For the parameter studies presented here, we use different resolutions,
because the higher gradient in density and electrical conductivity demand an
increase of both radial and horizontal resolutions.
For the Chebyshev polynomial truncations between $N_r\!=\!73$ and $N_r\!=\!145$ are
used while $\ell_{max}$ ranges between $85$ and $170$. Each simulation is run for
at least one magnetic diffusion time with the exception of some cases at the
lower Ekman number $E\!=\!10^{-5}$ (see Tab.~\ref{Tab2}).

Regarding the velocity boundary conditions, we apply a no-slip condition
at the inner core boundary and a free-slip condition at the outer boundary
in most of our simulations which seems appropriate for a gas planet with a rocky core.
A few test cases with no-slip conditions at both boundaries allow to
explore the impact of the boundary condition on the dynamics.

The existence of an inner core in Jupiter and its possible size
is still unclear. Here, we assume a small and electrically conducting solid
inner core with $r_i/r_o\!=\!0.2$.
GDW12 explore $\eta\!=\!0.2$ and $\eta\!=\!0.6$ in
very similar models with homogeneous electrical conductivity
and find generally very similar results.

Furthermore, we use constant entropy boundary conditions and match
the magnetic field to a diffusive solution at the inner boundary and
to a potential field at the outer boundary.
The Ekman number is either $E\!=\!10^{-4}$ or $E\!=\!10^{-5}$.
The larger value allows a more extensive scan of the
other system parameters like Rayleigh number, density stratification $N_{\rho}$
and electrical conductivity transition radius $\chi_m$.
At $E\!=\!10^{-5}$, we could only afford to run \tablecases{eleven} cases in a more
restricted parameter regime.
We assume a Prandtl number of $Pr\!=\!1$ and an inner boundary magnetic Prandtl
of typically $Pm_i\!=\!2$ for $E\!=\!10^{-4}$ and $Pm_i\!=\!1$ for
$E\!=\!10^{-5}$. \tablecases{Nine} additional cases
with $Pm_i\!=\!4-10$ at $E\!=\!10^{-4}$ and $Pm_i\!=\!3$ at
$E\!=\!10^{-5}$ have also been computed.

\begin{table}[h!]
\centering
\caption{\small{Values of critical Rayleigh number ($Ra_{cr}$) and critical wave number 
($m_{cr}$) for each $N_{\rho}$ \noc{at $\eta\!=\!0.2$}. The values were obtained with a 
modified version of the linear code by \cite{Jones09a}.}}
\begin{tabular}{cccc}
\hline
$N_{\rho}$ & $Ra_{cr}$ & $m_{cr}$ & $Ekman$ \\
\hline\hline
\vspace{-11pt}
 & & & \\
0.0 & $8.706\times 10^5$ & 4 & $10^{-4}$ \\
1.0 & $1.935\times 10^6$ & 5 & $10^{-4}$ \\
2.0 & $3.455\times 10^6$ & 6 & $10^{-4}$ \\
3.0 & $4.648\times 10^6$ & 43 & $10^{-4}$ \\
4.0 & $4.569\times 10^6$ & 49 & $10^{-4}$ \\
5.0 & $5.372\times 10^6$ & 55 & $10^{-4}$ \\
5.5 & $6.172\times 10^6$ & 58 & $10^{-4}$ \\
\hline
\vspace{-11pt}
 & & & \\
0.0 & $1.207\times 10^7$ & 7 & $10^{-5}$ \\
1.0 & $3.012\times 10^7$ & 9 & $10^{-5}$ \\
2.0 & $5.582\times 10^7$ & 11 & $10^{-5}$ \\
3.0 & $8.874\times 10^7$ & 108 & $10^{-5}$ \\
\hline
\end{tabular}
\label{Tab1}
\end{table}

All together, we ran \tablecases{74} cases with Rayleigh numbers between
$3$ and $46$ times supercritical.
\cite{Gastine12} examine anelastic convection for an aspect ratio of $0.6$ and
show that the critical Rayleigh number increases
with increasing stratification $N_{\rho}$. Table \ref{Tab1} demonstrates that
we observe a similar trend for the smaller aspect ratio $0.2$ employed here.
At a certain stratification, the critical wave number jumps from lower to high values.
This is the point where the centre of the flow convection moves from close to
the inner to close the outer boundary.
The respective transition
happens at larger stratifications $N_\rho$ when the Ekman number is
decreased \citep[see also][]{Jones09a}.

In our anelastic simulations, we consider a polytropic index of $m\!=\!2$ and
we explore density scale heights ranging from the Boussinesq
case $N_\rho\!=\!0$ to \tablecases{$N_\rho\!=\!5.5$}, where the latter corresponds to
a density jump of \tablecases{$\rho_i/\rho_o\!\simeq\!245$}.
While \textit{ab initio} simulations suggest a Jovian stratification of
$N_{\rho}\!=\!8.5$ from the bottom of the molecular
hydrogen layer to the
$1\,$bar level \citep{Guillot99,French12}. However, since the density
gradient rapidly steepens with radius in the planets outer shell
our largest stratifications already cover the %more than
inner 99\% of Jupiter's radius.

\subsection{Diagnostic parameters}
\label{parameters}

The parameters of all numerical experiments discussed here are listed in Tab.~\ref{Tab2} along
with several diagnostic quantities that characterize the solution and are defined in
the following.
The amplitude of the zonal flow contribution is measured in terms of
the Rossby number $Ro_{zon}$:
\begin{equation}
      Ro_{zon}=\frac{u_{zon}}{\Omega\,d}
      \textrm{,}\;\;\textrm{ with }\;\;
      u_{zon}=\sqrt{\frac{3}{r_o^3-r_i^3}\int_{r_i}^{r_o}\!\langle \overline{u}_\phi^2\rangle\,r^2\,
            \mathrm{d}r}
      \,\textrm{,}
\label{eq:rozon}
\end{equation}
where \noc{$u_{zon}$ is the rms volume-averaged flow velocity and} the triangular brackets
denote the angular average
\begin{equation}
      \big\langle f\big\rangle=\frac{1}{4\pi} \,
            \int_{0}^{\pi}\! \int_{0}^{2\pi}\!
            f(r,\theta,\phi) \sin\!\theta \, \mathrm{d}\theta \, \mathrm{d}\phi
      \textrm{,}
\label{eq:volave}
\end{equation}
$\overline{u}_\phi$ is the axisymmetric azimuthal flow component,
and $V$ is the volume of the spherical shell.
Overbars correspond to azimuthal averages.
We use the relative kinetic energy
\begin{equation}
      Z=\frac{Ro_{zon}^2}{Ro^2} =
         \frac{\int_{r_i}^{r_o}\big\langle \overline{u}_\phi^2\big\rangle\,\mathrm{d}r}
             {\int_{r_i}^{r_o}\big\langle u^2\big\rangle\,\mathrm{d}r}
\label{eq:Z}
\end{equation}
to quantify the relative importance of zonal flows.

The magnetic Reynolds number $R_m$ estimates the
ratio of magnetic field production and diffusion and we use a modified form
here to account for the radial-dependent magnetic diffusivity:
\begin{equation}
      R_m=\frac{3}{r_o^3-r_i^3} \int_{r_i}^{r_o}\!
            \frac{\sqrt{\big\langle u^2(r,\theta,\phi)\big\rangle}}{\tilde{\lambda}(r)}
            \,r^2\, \mathrm{d}r
      \textrm{.}
\label{eq:rm}
\end{equation}

The local Rossby number has been introduced by \cite{Christensen06}
to quantify the relative importance of the advection term
in the Navier-Stokes equation (Eq.~\ref{eq:navierstokeseq}) and
is defined as
\begin{equation}
      Ro_\ell=
            \frac{\sqrt{\frac{1}{V}\,\int_{r_i}^{r_o}\! \big\langle u^2 \big\rangle
            \,r^2\, \mathrm{d}r}}{\Omega\,\ell}
      \textrm{.}
\label{eq:rol}
\end{equation}
Here, $\ell$ is a typical flow length scale given by
\begin{equation}
      \ell(r)=\frac{\pi \, u^2(r)}{\displaystyle\sum\limits_l l \, u_l^2(r)}
\label{eq:ell}
\end{equation}
where $u_l$ is the flow contribution of spherical harmonic degree $l$.
We use a modified form of $Ro_\ell$ based exclusively on the inner conducting region
($r_i\le r\le r_m$):
\begin{equation}
      Ro_\ell=\frac{3}{r_o^3-r_i^3} \int_{r_i}^{r_m}\!
            \frac{\sqrt{\big\langle u^2(r,\theta,\phi)\big\rangle}}{\Omega\,\ell(r)}
            \,r^2\, \mathrm{d}r
      \textrm{,}
\label{eq:rol2}
\end{equation}

The magnetic field strength is quantified by the Elsasser number which measures the ratio
of Lorentz to Coriolis forces using the modified form
\begin{equation}
      \Lambda=\frac{3}{\mu_0\,\Omega\,(r_o^3-r_i^3)} \,
            \int_{r_i}^{r_o} \Bigg\langle\frac{\mathbf{B}^2}{\rho(r)\,\tilde{\lambda}(r)}\Bigg\rangle
            \, r^2 \, \mathrm{d}r
      \textrm{.}
\label{eq:elsasser}
\end{equation}
The geometry of the surface field is characterized by the dipolarity
\begin{equation}
      f_{dip}=\frac{\Big\langle \big({\mathbf{B}_{l=1}^{m=0}}\big)^2 \Big\rangle}
            {\Bigg\langle \displaystyle\sum\limits_{l,m\le 12} \big({\mathbf{B}_l^m}\big)^2
            \Bigg\rangle}
      \textrm{,}
\label{eq:dip}
\end{equation}
which measures the relative energy in the axial dipole contribution at the outer boundary $r_o$.
Following \cite{Christensen06}, we restricted the magnetic field to spherical harmonic
degrees and orders below 12 in Eq.~(\ref{eq:dip}).
Tab.~\ref{Tab2} lists time averages of the properties defined above
for all our models and we always refer to the time-averaged
properties for characterizing our solutions in the following.
The time variability of the dipolarity, also listed
in Tab.~\ref{Tab2}, is quantified by its standard deviation $S\!D_{dip}$.

\section{Results}
\label{results}

\subsection{Dynamo regimes}
\label{rol}

\begin{figure}[h!]
\begin{center}
{\centering \includegraphics[trim=1cm 1cm 0.5cm 0cm, height=3.2in]{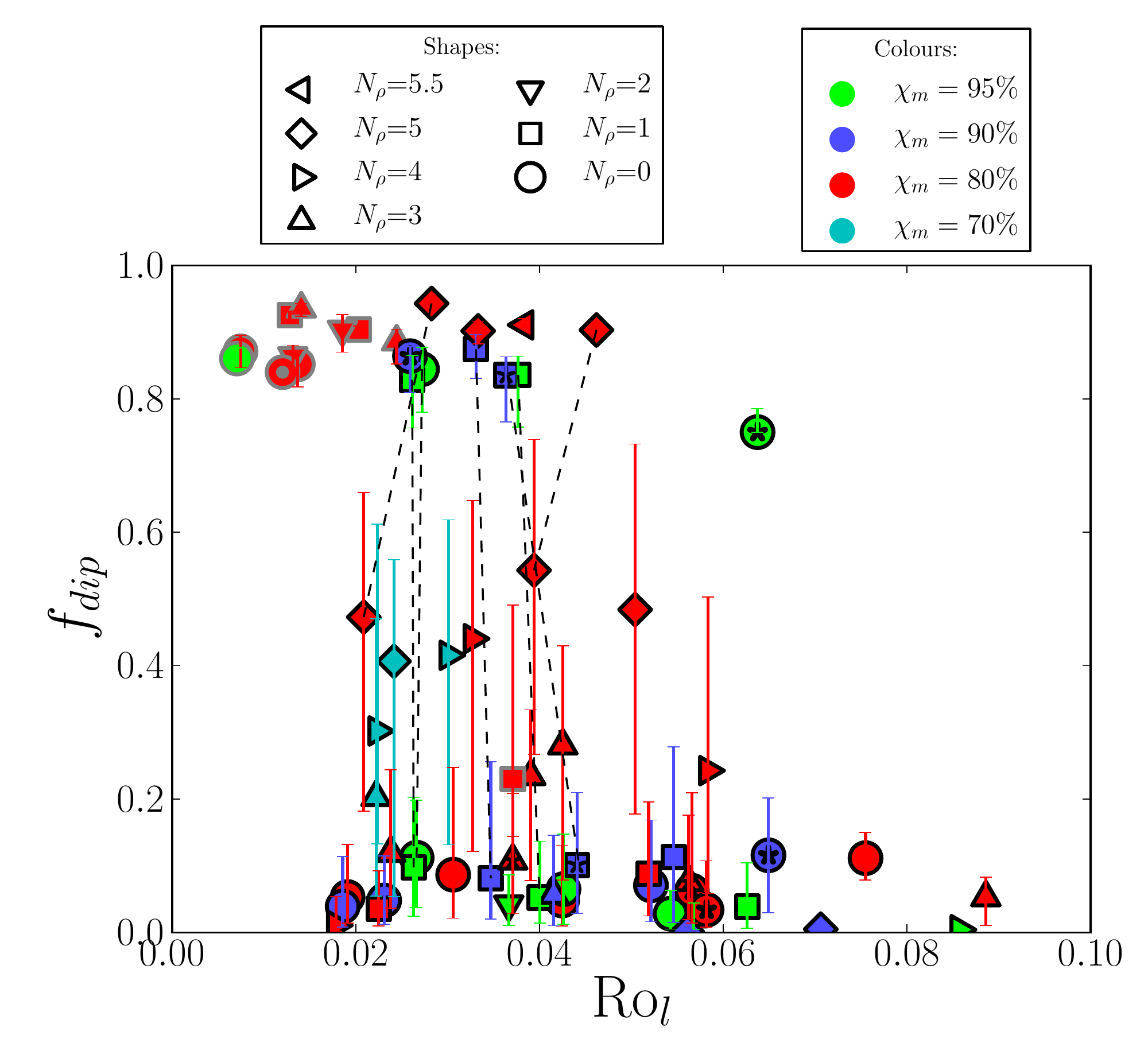} }
\caption{\small Dipolarity against the local Rossby number defined by Eq.~(\ref{eq:rol2}).
The outer line of each symbol represents the Ekman number:
black -- $E\!=\!10^{-4}$/$Pm=2$ and grey -- $E\!=\!10^{-5}$/$Pm=1$.
The black star inside the symbols marks the cases with a no-slip upper boundary,
instead of free-slip.
The error bars correspond to standard deviations of the time series of each
case, for which the point itself is the time average listed in Tab.~\ref{Tab2}.
The \tablecases{seven} dashed lines connect
\tablecases{seven} sets of cases for which we found two solutions,
depending on the initial magnetic field.
The Boussinesq case with a grey dot inside is the case
from \cite{Heimpel11} of $\chi_m\!=\!80\%$ and $\eta\!=\!0.35$.\label{fig:DipvsRol}}
\end{center}
\end{figure}

\begin{figure}[h!]
\begin{center}
{\centering \includegraphics[trim=1cm 1cm 0.5cm 0.3cm, height=3.2in]{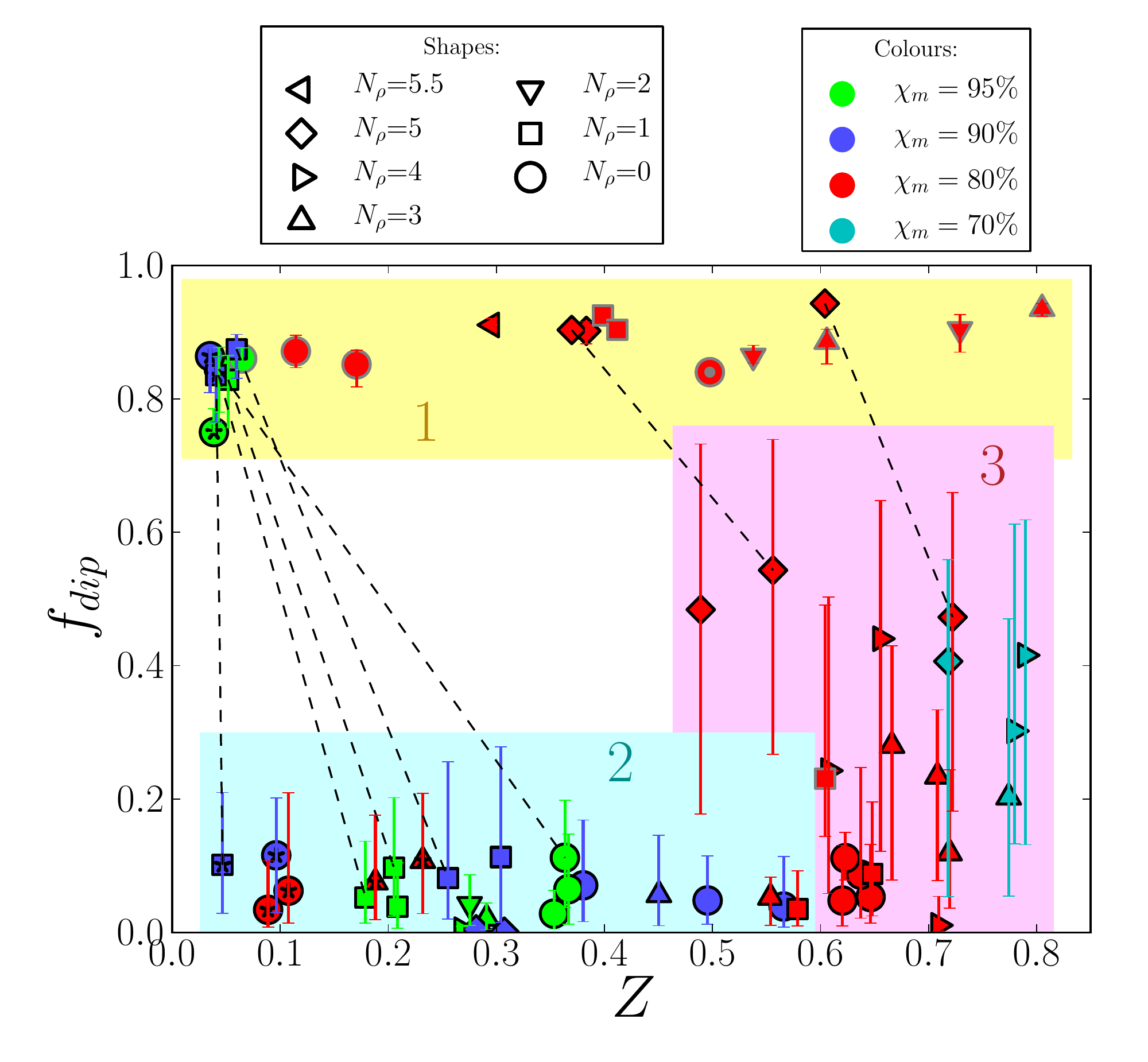} }
\caption{\small Dipolarity plotted against the ratio between the azimuthal
kinetic energy and the total kinetic energy, averaged in time
and volume. The symbols and colours have the same definition as in Fig.~\ref{fig:DipvsRol}.
The three boxes mark the three different regimes discussed in the text.
The \tablecases{seven} dashed lines connect
\tablecases{seven} sets of cases for which we found two solutions, depending on the initial
magnetic field.\label{fig:DipvsEkin}}
\end{center}
\end{figure}

In the complex models explored here, the magnetic field geometry not only
depends on the local Rossby number \citep{Christensen06} but also on the density
stratification, on the thickness of the weaker conducting layer,
on the Ekman number, and on the magnetic Prandtl number.
Fig.~\ref{fig:DipvsRol} shows the dependence of
the dipolarity $f_{dip}$ on the local Rossby number for all our cases,
excluding runs with $Pm\!>\!2$, with the exception of case \tablecases{67} from
Tab.~\ref{Tab2} which is similar to one of \cite{Heimpel11}'s runs.
To illustrate the relation between the field geometry and the
zonal flows, we plot $f_{dip}$ versus the relative kinetic energy of
axisymmetric azimuthal flows in Fig.~\ref{fig:DipvsEkin}.
In both figures, the symbol type refers to the different stratifications while
the symbol colour identifies the four transitional radii $\chi_m$ explored here.
We start by analysing the different dynamo regimes
based on the results for $E\!=\!10^{-4}$ and $Pm\!=\!2$ and come back to the
solutions for larger magnetic Prandtl numbers and for $E\!=\!10^{-5}$ further
below.

When the weakly conducting layer is relatively thin
($\chi_m\!=\!95\%$ and $\chi_m\!=\!90\%$)
and the stratification is mild to intermediate $(N_\rho\!\le\!2)$, we find two
distinct branches. A dipolar branch, characterized by $f_{dip}\!>\!0.7$ and
weak zonal flows, is restricted to cases with local Rossby
numbers below the critical value of $Ro_{\ell c}\approx0.04$.
This is significantly lower than the values of $Ro_{\ell c}\approx 0.08$
suggested for homogeneous electrical conductivity by GDW12.
The dipole-dominated solutions forming this branch are located
in the upper left corner of Fig.~\ref{fig:DipvsRol} and in the
left portion of the yellow high-dipolarity regime in
Fig.~\ref{fig:DipvsEkin}.

A second branch with multipolar magnetic fields at $f_{dip}\!<\!0.2$
but intermediate zonal flows exists for all $Ro_\ell$ values.
These solutions can be found in the lower part of
Fig.~\ref{fig:DipvsRol} and the cyan low-dipolarity regime in
Fig.~\ref{fig:DipvsEkin}.

For local Rossby numbers below $Ro_{\ell c}\approx 0.04$,
we thus find both types of solutions while only multipolar
solutions remain stable beyond $Ro_{\ell c}$.
Figures~\ref{fig:DipvsRol} and \ref{fig:DipvsEkin}
contain \tablecases{seven} examples (dashed lines) where a solution on each
branch is found for identical model parameters, clearly demonstrating
the bistability for $Ro_\ell<Ro_{\ell c}$. Which branch a specific
numerical simulation will chose depends on the initial magnetic
field configuration. Note that the multipolar attractor always
has the more intense zonal flows (see Fig.~\ref{fig:DipvsEkin}).
Comparing magnetic Reynolds numbers and local Rossby numbers for
bistable cases shows that the relative difference is smaller in the
latter than in the former measure. This indicates that the weaker flow
amplitude caused by the larger Lorentz forces in the dipole-dominated cases
is accompanied by a growth in the flow length scale.

Increasing the stratification to values beyond $N_\rho\!=\!2$ while
keeping $\chi_m$ large always leads to solutions of
the multipolar thin-shell type discussed by GDW12.
Altogether, the behaviour for a thin weakly conducting layer is
similar to that for a homogeneous electrical conductivity with the
exception of the lower critical Rossby number $Ro_{\ell c}$.

\begin{figure}[h!]
\begin{center}
{\centering\includegraphics[trim=0cm 0.5cm 0.5cm 0cm, height=2.15in]{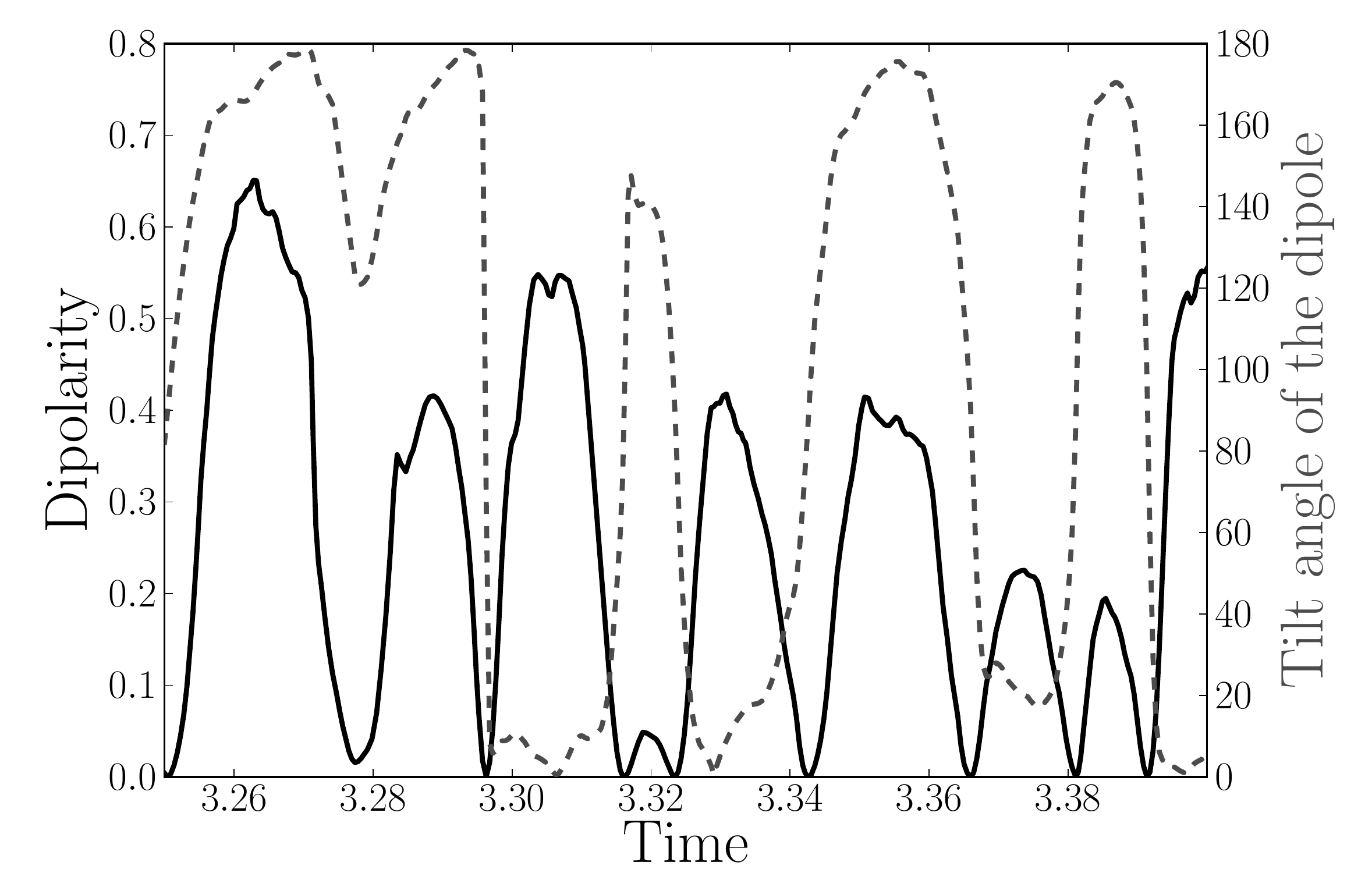} }
\caption{\small Time evolution of the dipolarity at the surface (solid black line) and the tilt
angle of the dipole (dashed grey line) in degrees. The time is given in magnetic diffusion units.
The parameters are:
$E\!=\!10^{-4}$, $N_{\rho}\!=\!3$, $Ra\!=\!4.3\,Ra_{cr}$, $\chi_m\!=\!80\%$
(case \tablecases{33} from Tab.~\ref{Tab2}).
\label{fig:diptilt}}
\end{center}
\end{figure}

For a thicker weakly conducting layer with $\chi_m\!=\!80$\%, the
influence of the stratification on the dipolarity is reversed.
Clearly, dipolar solutions with $f_{dip}\!>\!0.7$ now exclusively
exist for stratifications of $N_\rho\!=\!5$ or $N_\rho\!=\!5.5$.
Since the relative zonal flow amplitude reaches intermediate values,
these cases can be found in the middle section of the yellow
regime in Fig.~\ref{fig:DipvsEkin}.
A second branch of solutions is characterized by low to intermediate dipolarity
that increases with $N_\rho$ and by large relative zonal flow amplitudes.
These cases populate the pink region in Fig.~\ref{fig:DipvsEkin}.
For stratification of $N_\rho\!\ge\!3$, the solutions on this secondary
branch become strongly time-dependent as indicated by the large error
bars in Figs.~\ref{fig:DipvsRol} and \ref{fig:DipvsEkin}.
\Cref{fig:diptilt} demonstrates that the time dependence reflects
an oscillation between dipolar and multipolar field configurations
without ever establishing a solution on the dipole-dominated
branch. Polarity reversals or excursions become possible
when the dipolarity is relatively low.

Once more, both branches coexist for not too large local
Rossby numbers and we could identify two bistable cases
for $\chi_m\!=\!80$\%, $N_\rho\!=\!5$ and $Ra/Ra_{cr}\!=\!7.4$, $Ra/Ra_{cr}\!=\!9.3$.
When increasing the Rayleigh number to $Ra/Ra_{cr}\!=\!11.2$, however,
only the multipolar solution remains which suggests a critical local
Rossby number of about $Ro_{\ell c}\!\approx\!0.5$ (see Tab.~\ref{Tab2}).

For $\chi_m\!=\!70$\%, the thickest weakly conducting outer shell explored here,
even stronger stratification seems required to establish a dipole dominated
magnetic field than at $\chi_m\!=\!80$\%.
For the magnetic Prandtl number $Pm\!=\!2$, generally used at
$E\!=\!10^{-4}$, only the highly time-dependent solutions with
intermediate dipolarity and strong zonal winds (on average) were found,
even at $N_\rho\!=\!5$. However, the mean dipolarity increases with $N_\rho$,
just as in the $\chi_m\!=\!80$\% cases and stratifications of $N_\rho\!>\!5$
may finally establish a dipole-dominated solution.

\begin{figure}[h!]
\begin{center}
{\centering \includegraphics[trim=0cm 1cm 0.5cm 0cm, height=2.48in]{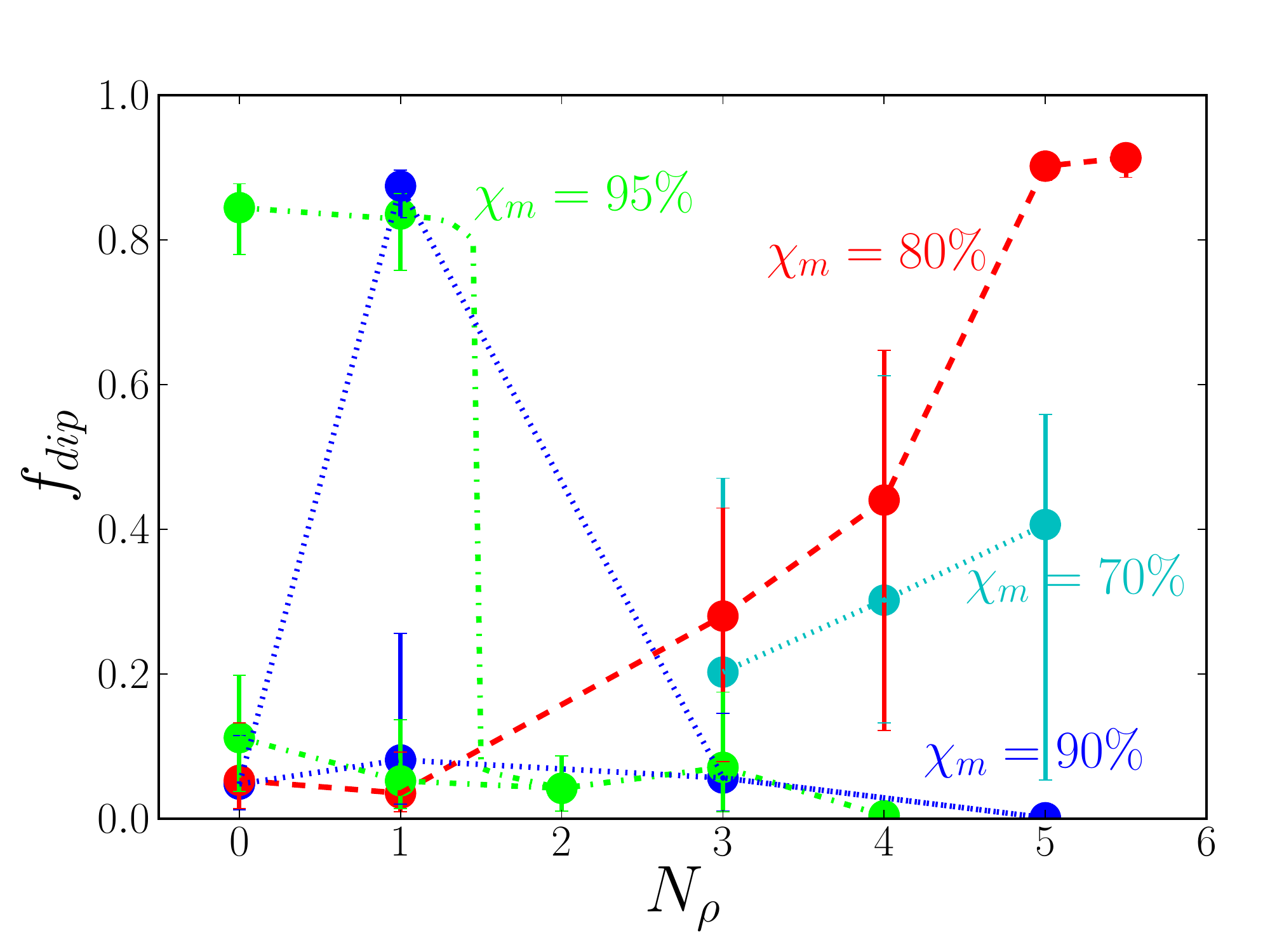} }
\caption{\small Dipolarity against density stratification, for fixed
transition radii: green -- $\chi_m=95\%$, blue -- $\chi_m=90\%$, red -- $\chi_m=80\%$,
cyan -- $\chi_m=70\%$.
The dashed lines simply gather the symbols with the same $\chi_m$.
The error bars are the same as in Figs.~\ref{fig:DipvsRol} and \ref{fig:DipvsEkin}.
\label{fig:DipvsNrho}}
\end{center}
\end{figure}

\Cref{fig:DipvsNrho} highlights the role of the density stratification
at $\chi_m\!=\!95$\%, $90$\%, $80$\% and $70$\% for models at
$E\!=\!10^{-4}$ and $Pm\!=\!2$ with similar local Rossby numbers.
When the weakly conducting layer is relatively thin ($\chi_m\!=\!95\%$ or $\chi_m\!=\!90\%$),
dipole-dominated solutions can only be found for $N_\rho\!<\!2$ while multipolar
solutions exist for all stratifications.
For a thicker layer, however, the stratification
has a reversed effect. The mean dipolarity
increases with stratification and highly dipolar solutions are
only found for stronger stratifications $N_\rho\!\ge\!5$.
Note that at $\chi_m\!=\!90$\% and $N_\rho\!=\!0$,
we could only find multipolar solutions, even at low Rayleigh numbers
where $Ro_\ell$ is small. The reason for this is not yet understood.
What finally helped to establish a dipolar solution here was increasing
the magnetic Prandtl number from $2$ to $5$.

We also tested the effect of larger magnetic Prandtl numbers for several other
parameter combinations and this often promoted dipole-dominated solutions.
For example, at $\chi_m\!=\!80$\% and $N_\rho\!=\!0$ a multipolar case became
dipolar when increasing $Pm$ from $2$ to $10$. Likewise, the highly
time-dependent case at $\chi_m\!=\!80$\%, $N_\rho\!=\!4$, $Ra/Ra_{cr}\!=\!5.5$ and $Pm\!=\!2$
developed into a stable dipole-dominated solution when doubling $Pm$.
The same behaviour was found at $\chi_m\!=\!70$\%, $N_\rho\!=\!4$ and $Ra/Ra_{cr}\!=\!6.6$.
This indicates a certain trade-off between larger stratifications and
higher electrical conductivities.
At $\chi_m\!=\!80$\% or $70$\%, $N_\rho\!=\!3$ and $Ra/Ra_{cr}\!=\!4.3$, however, an increase
from $Pm\!=\!2$ to $Pm\!=\!6$ was not sufficient to establish a dipole-dominated
solution. Even higher magnetic Prandtl numbers may be required here.

\begin{figure*}[ht]
\begin{center}
{\centering
      \includegraphics[trim=0cm 1cm 0cm 0cm, height=3.2in]{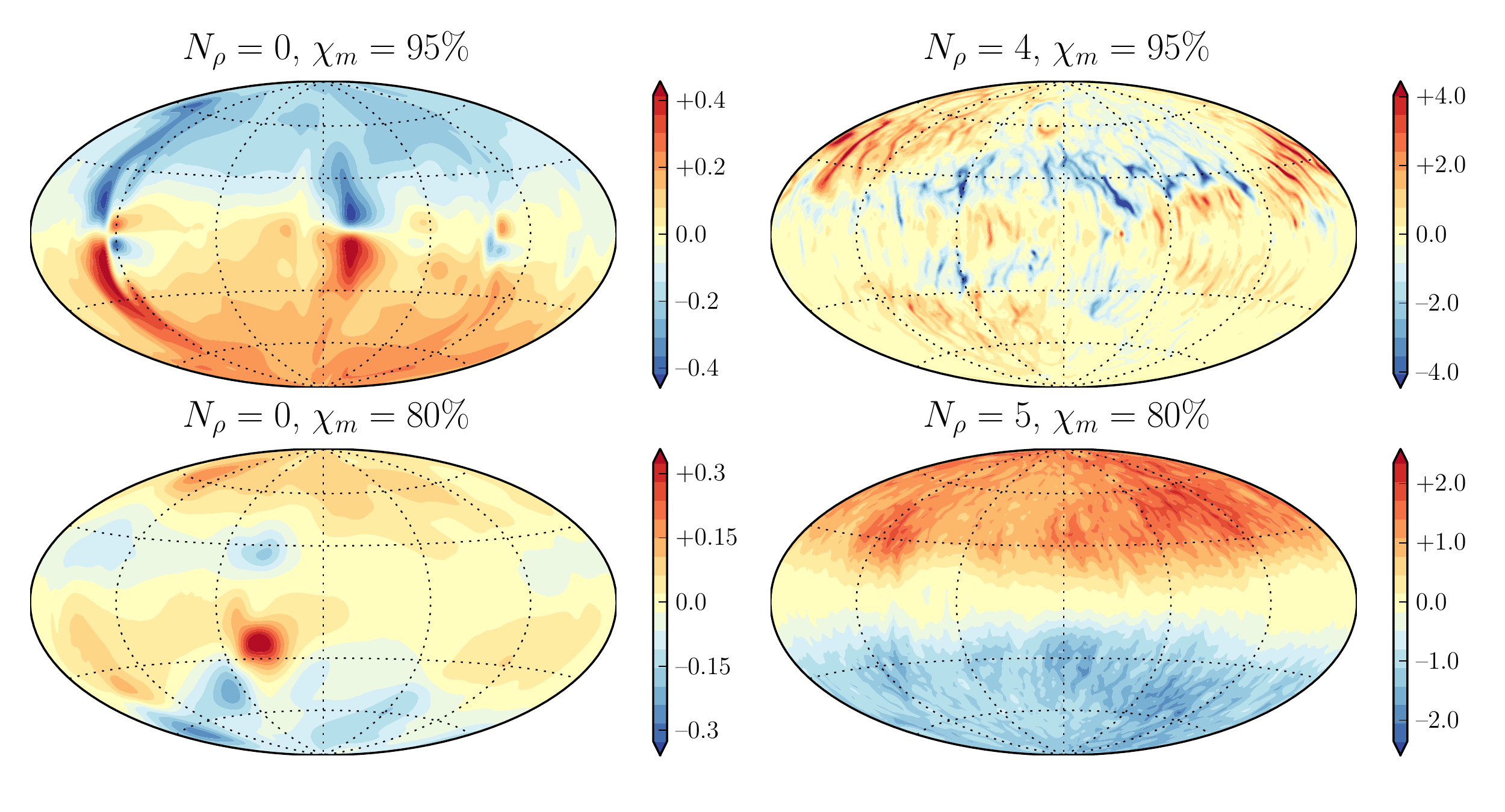}
}
\caption{\small Radial magnetic field at the outer boundary.
The top row corresponds to $\chi_m\!=\!95\%$ (cases \tablecases{1d} and \tablecases{40}
of Tab.~\ref{Tab2}, respectively) and the bottom row to
$\chi_m\!=\!80\%$ (cases \tablecases{16} and \tablecases{51} of the same table).
The maps on the left
are $N_{\rho}\!=\!0$ cases and the maps on the right column are $N_{\rho}\!=\!4$
(top) and $N_{\rho}\!=\!5$ (bottom). Magnetic fields are given in units of Elsasser
number.\label{fig:brs}}
\end{center}
\end{figure*}

\Cref{fig:brs} illustrates the different types of solutions with
snapshots of the radial magnetic field at the outer
boundary for $E\!=\!10^{-4}$.
The top row shows $\chi_m\!=\!95\%$ cases at two different stratifications:
a dipole-dominated Boussinesq case and a multipolar solution
at $N_{\rho}\!=\!4$.
The latter shows a large scale wave number ($m\!=\!1$) structure similar to
that reported for multipolar dynamos with homogeneous electrical
conductivity and free-slip boundaries \citep[][GDW12]{Goudard08}.
The bottom row of Fig.~\ref{fig:brs} depicts the two
branches found for $\chi_m\!=\!80\%$.
The left panel shows a snapshot of a Boussinesq multipolar
case and the right panel illustrates the dipolar configuration
found at strong stratifications ($N_{\rho}\!=\!5$).

The results by \cite{Heimpel11} prompted us to also conduct simulations
at the lower Ekman number of $10^{-5}$ used in their study.
Tab.~\ref{Tab2} lists the respective models with different
Rayleigh numbers and stratifications. The thicker weakly
conducting layer of $\chi_m\!=\!80$\% was generally chosen except
for one model with $\chi_m\!=\!95$\%.
The Boussinesq case \tablecases{67} is identical to one of the models
presented by \cite{Heimpel11} and has a larger aspect ratio of
$\eta\!=\!0.35$ instead of $\eta\!=\!0.2$.
The outer grey line of the symbols in Figs.~\ref{fig:DipvsRol} and \ref{fig:DipvsEkin}
corresponds to the \tablecases{twelve} runs with $E\!=\!10^{-5}$.

While for $E\!=\!10^{-4}$ and $\chi_m\!=\!80$\% we had to increase the
stratification to $N_\rho\!\ge\!5$ to find strongly dipolar solutions, this
is not the case any more at $E=10^{-5}$.
Even the Boussinesq models now clearly have dipole-dominated magnetic fields.
The only multipolar case, which seems to be of the highly time-dependent
type, is found at $N_\rho\!=\!1$ and it has a local Rossby number of
$Ro_{\ell}\!\approx\!0.04$. Another model at $N_\rho\!=\!1$ but
$Ro_{\ell}\!\approx\!0.02$ is strongly dipolar so that the
critical local Rossby number can be estimated to $Ro_{\ell c}\!\approx\!0.03$.
Note however, that $Ro_{\ell c}$ may depend on stratification.
%which helped to promote dipolar solutions at $E=10^{-4}$.

\subsection{The role of zonal flows}
\label{diporflow}

The coexistence of dipolar and multipolar branches indicates a competition
between zonal winds and dipolar magnetic fields already discussed by GDW12.
The stronger Lorentz forces associated to the larger dipolar fields
effectively compete with the Reynolds stresses responsible for
driving the zonal winds. The zonal wind amplitude and the relative
zonal wind energy thus remain typically small.
Fierce zonal winds, on the other hand, seem to promote multipolar fields.
This is at least the situation for $E\!=\!10^{-4}$ and $\chi_m\!=\!95$\% or
$90$\%. But why are stronger stratifications and/or larger
magnetic Prandtl numbers required to yield dipole-dominated dynamo
action for thicker weakly conducting outer layers?

\begin{figure*}[ht]
\begin{center}
{\centering
      \hspace{10pt}\\
      \includegraphics[trim=0cm 1cm 0cm 0cm, height=5.5in]{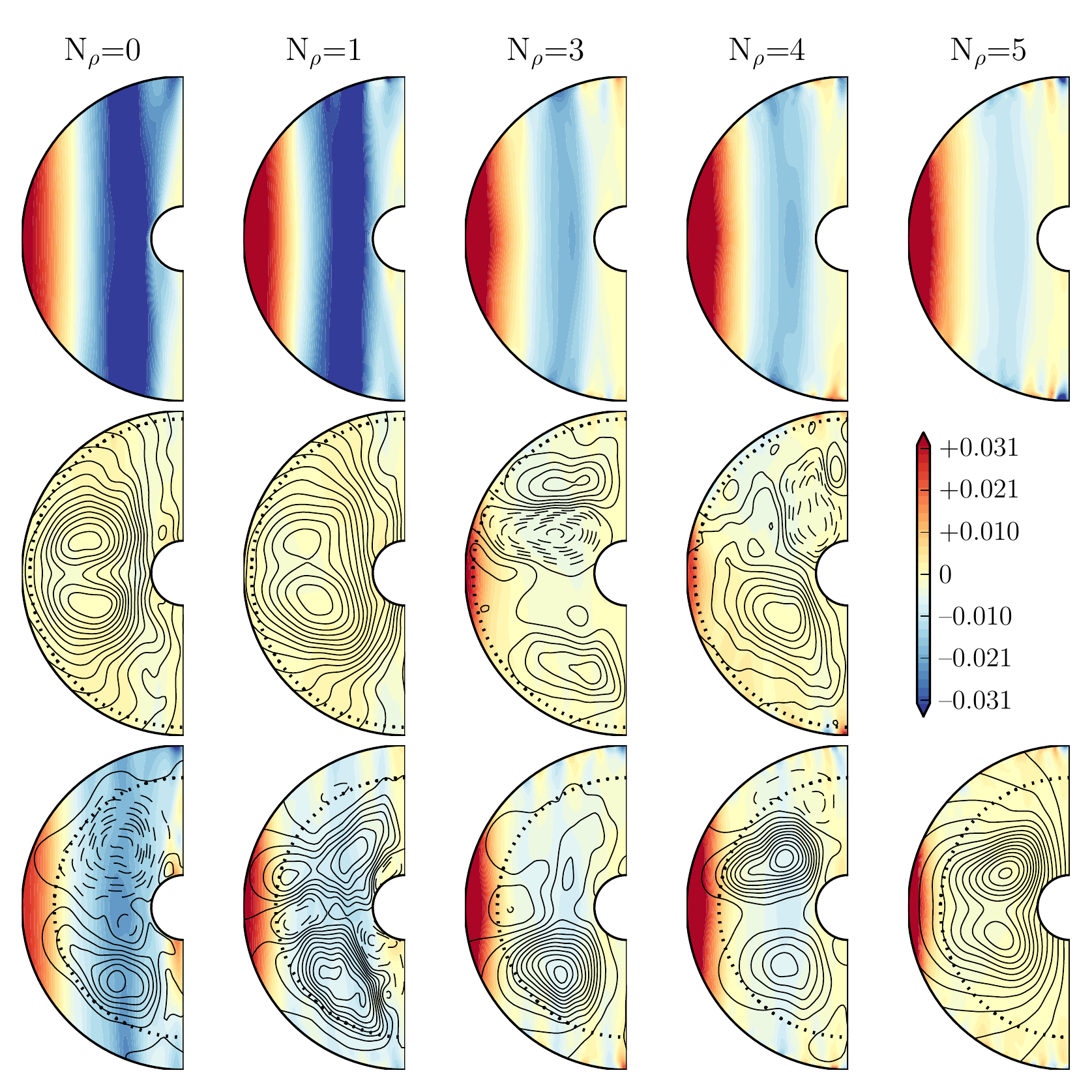}\\
      \hspace{7pt} }
\caption{\small Azimuthal averages of the zonal component of the flow.
Each column of three plots has a different $N_{\rho}$, namely 0, 1, 3, 4 and 5 from left to right.
In the bottom and middle rows, the poloidal field lines are plotted on top of
the zonal velocity contours in units of Rossby number $Ro=u/(\Omega r_o)$.
The dotted line in the middle and bottom rows corresponds to
$r_m\!=\!95\%$ and $r_m\!=\!80\%$, respectively.
The top row shows the corresponding hydrodynamical solutions.
\label{fig:zonalflow}}
\end{center}
\end{figure*}

\Cref{fig:zonalflow} illustrates the zonal flow structure and the poloidal
magnetic field lines for different stratifications at $\chi_m\!=\!95$\% (middle row)
and $\chi_m\!=\!80$\% (bottom row) for $E\!=\!10^{-4}$.
The top row shows non-magnetic cases and
demonstrates that the inner retrograde jet decreases in amplitude when
the stratification intensifies. This reflects the progressive outward
concentration of the convective motions and thus of the Reynolds stresses
driving the zonal flows (GDW12).

The dominance of Coriolis forces at this
relatively low Ekman number enforces the Taylor-Proudman theorem and
the intense zonal jets remain strongly geostrophic, i.e.~variations in the direction
of the rotation axis are much smaller than variations perpendicular
to it.
For the thinner weakly conducting layer (middle row in Fig.~\ref{fig:zonalflow}),
the Lorentz forces associated
with the stronger dipolar field at mild stratifications effectively
suppress the zonal flows in the whole shell.
For $N_\rho\!>\!2$, the weaker multipolar fields created by the thin-shell dynamo
allow the outer prograde jet to survive,
albeit with a significantly reduced amplitude and a restricted width
than in the non-magnetic simulations.
The thickness of the weakly conducting layer now determines the width
of the outer jet, confirming previous work by \cite{Heimpel11}.

At $\chi_m\!=\!80$\% (lower row in Fig.~\ref{fig:zonalflow}), the zonal flows
generally remain more energetic than for $\chi_m\!=\!95$\%. Since these flows
are largely geostrophic, the force balances on geostrophic cylinders
(i.e.~on cylinders aligned with the rotation axis) should be considered.
The Lorentz forces now have a harder time to brake the zonal flows since
they act in a significantly reduced volume. Dipole-dominated dynamo action
only becomes possible when the retrograde inner zonal jet is already relatively weak in
the non-magnetic simulations, which happens at stronger stratifications. The thin-shell dynamo
mechanism generating the multipolar field for $\chi_m\!\ge\!90$\% does not apply here,
since it would have to operate, at least partly, in the weakly conducting layer
where the magnetic Reynolds number is now too low to support dynamo action.
Instead, a strongly dipolar magnetic field is generated in the deeper
interior where it does not interfere with the remaining prograde
outer zonal jet.

\begin{figure}[h!]
\begin{center}
{\centering
      \includegraphics[trim=0cm 1cm 0.5cm 0cm, height=2.45in]{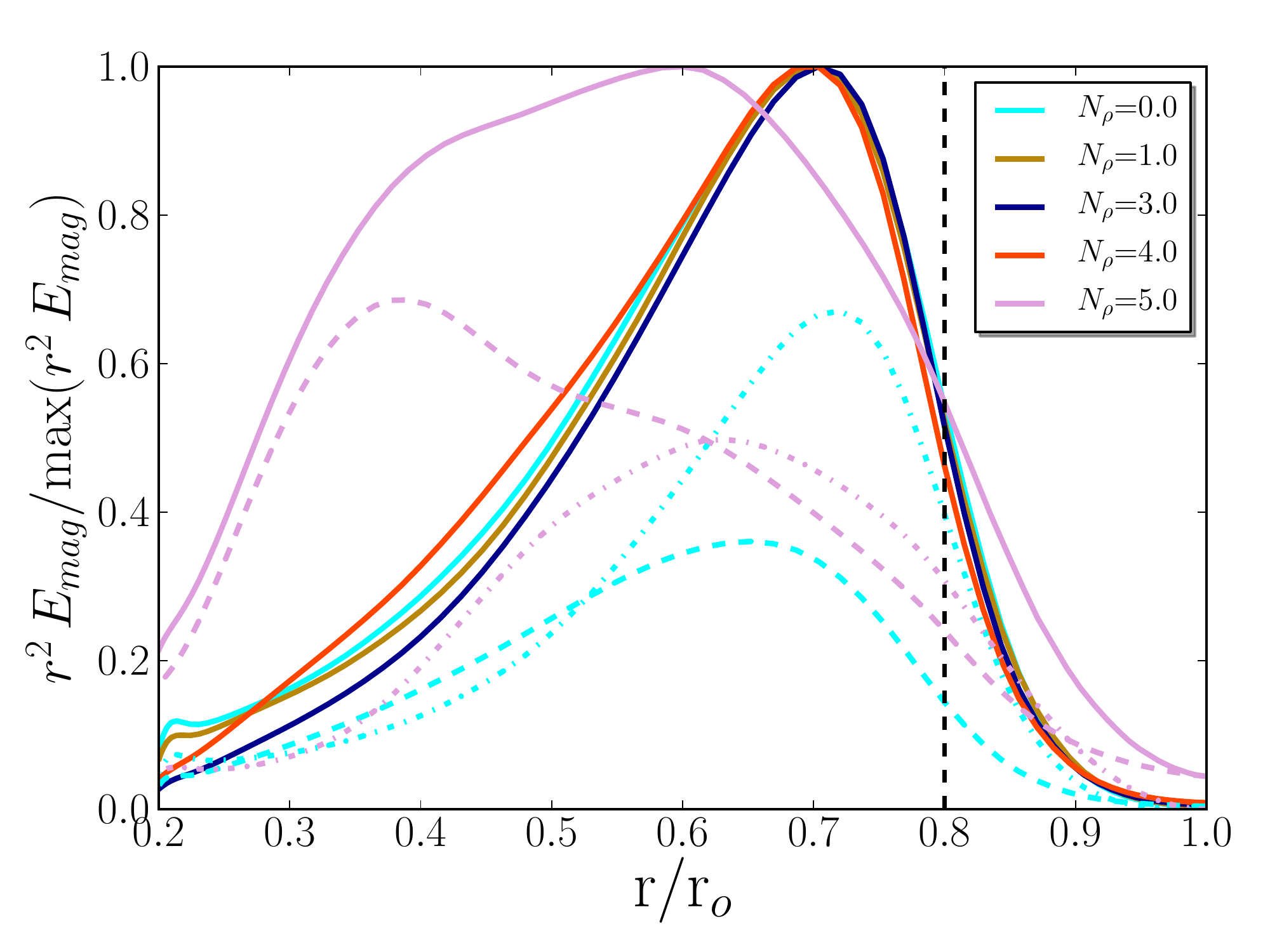}
      }
\caption{\small Radial profile of magnetic energy flux ($r^2\,E_{mag}$) averaged over time.
The dashed black line is the location of $\chi_m\!=\!80\%$. These results
correspond to the red triangles and red dashed line from Fig.~\ref{fig:DipvsNrho}.
The poloidal (dashed lines) and toroidal (dot-dashed lines) components are
also shown for $N_{\rho}\!=\!5$ and $N_{\rho}\!=\!0$, with the corresponding colours.
The magnetic energy fluxes are normalized by their maximum values.
\label{fig:EMagvsRad}}
\end{center}
\end{figure}

The change in the depth of the poloidal dynamo action is further illustrated
by the radial profiles of magnetic energy shown in Fig.~\ref{fig:EMagvsRad}.
Up to a stratification of $N_\rho\!=\!4$, poloidal and toroidal magnetic
energies have similar profiles and peak in the outer part of the
conducting region around $r/r_o\!\simeq\!0.7$.
For $N_\rho\!=\!5$, however, the profiles are different with
a pronounced focus on deeper parts of the shell around $r/r_o\!=\!0.4-0.6$.
For $N_\rho\!\le\!4$, the toroidal magnetic field is larger than the poloidal,
suggesting that the induction mechanism is different from the dipole-dominated
case at $N_\rho\!=\!5$ where the poloidal field is stronger.

At $\chi_m\!=\!70$\%, the volume over which Lorentz forces can
efficiently brake zonal winds is even further reduced.
Dipolar solutions can only be found at even larger stratification
than at $\chi_m\!=\!80$\% where the driving of the inner zonal jet is
yet weaker.

The role of Lorentz forces in defeating zonal winds and thereby
enabling dipole-dominated magnetic fields also offers an explanation
why larger magnetic Prandtl numbers help.
The reason likely is that larger $Pm$ values lead to stronger
magnetic fields and thus stronger Lorentz forces.
We can also now interpret the highly time-dependent solutions with intermediate
mean dipolarities. Here, the balance seems to be undecided (Fig.~\ref{fig:diptilt}).
Stronger Lorentz forces successfully suppress the zonal winds at times but
never enough to establish the solution on the highly dipolar more stable branch.
At other times, Reynolds stresses succeed in driving stronger zonal flows that
mostly create a weaker multipolar magnetic field.

%RIGID BOUNDARY:\\
To further test the theory that the zonal flows are decisive for the
field geometry we ran a few $E\!=\!10^{-4}$ cases with a no-slip outer
boundary condition that largely prevents zonal flows from developing.
The results are mixed and not entirely conclusive, which may have to do with the
fact that other flow components are also affected by this change in boundary
conditions.
At $\chi_m\!=\!95$\%, $N_\rho\!=\!0$ and $Ra/Ra_{cr}\!=\!23.0$, the no-slip
boundary conditions indeed promote a dipole-dominated solution with weak
zonal flows where we only find multipolar solutions with strong zonal
flows for a free-slip outer boundary condition
(compare cases \tablecases{3} and \tablecases{$4$}).
The same positive effect was found for  $\chi_m\!=\!90$\%, $N_\rho\!=\!0$
and $Ra/Ra_{cr}\!=\!11.5$
(cases \tablecases{7} and \tablecases{$8$}).
At $\chi_m\!=\!90$\%, $N_\rho\!=\!1$ and $Ra/Ra_{cr}\!=\!5.2$, however, we
find bistable cases for both type of boundary conditions (cases \tablecases{22d/m}
and \tablecases{$23$d/m}).
In the no-slip case, both the dipole-dominated and the multipolar
solution have weak zonal flows. Free-slip outer boundary condition
promotes dipolarity, but it is not a necessary condition to find this
feature. Note that such a bistable case for no-slip conditions has already
been reported by \cite{Christensen06}.

At $\chi_m\!=\!80$\%, $N_\rho\!=\!0$ and $Ra/Ra_{cr}\!=\!23.0$, the suppression
of the zonal flows by the no-slip condition is not sufficient to
yield a dipole-dominated solution and the same is true
at $\chi_m\!=\!80$\%, $N_\rho\!=\!3$ and $Ra/Ra_{cr}\!=\!3.2$ or $Ra/Ra_{cr}\!=\!4.3$
(cases \tablecases{17}, \tablecases{32} and \tablecases{36}).
In the latter two examples, the particular thin shell
dynamo described by GDW12, rather than the stronger zonal flows,
may be the reason for the multipolarity which could explain why the
no-slip condition has no effect.

%HERE POTENTIALLY THE NO-LORENTZ-FORCE CASES

%HIGH a:\\
We also varied the electrical conductivity profile in a few cases.
Increasing the exponential decay rate from $a\!=\!9$ to $a\!=\!25$
for two simulations at $\chi_m\!=\!80$\% required a finer radial numerical grid
and thus more expensive numerical simulations.
The zonal flows in the weakly conducting layer were intensified in both
cases, likely because of the further decreased weaker Lorentz forces there.
The type of solution, however, remained unchanged
(see cases \tablecases{33/34} and \tablecases{51/52}).

\begin{figure}[h!]
\begin{center}
{\centering
      \includegraphics[trim=0.5cm 0cm 1cm 0cm, height=3.2in]{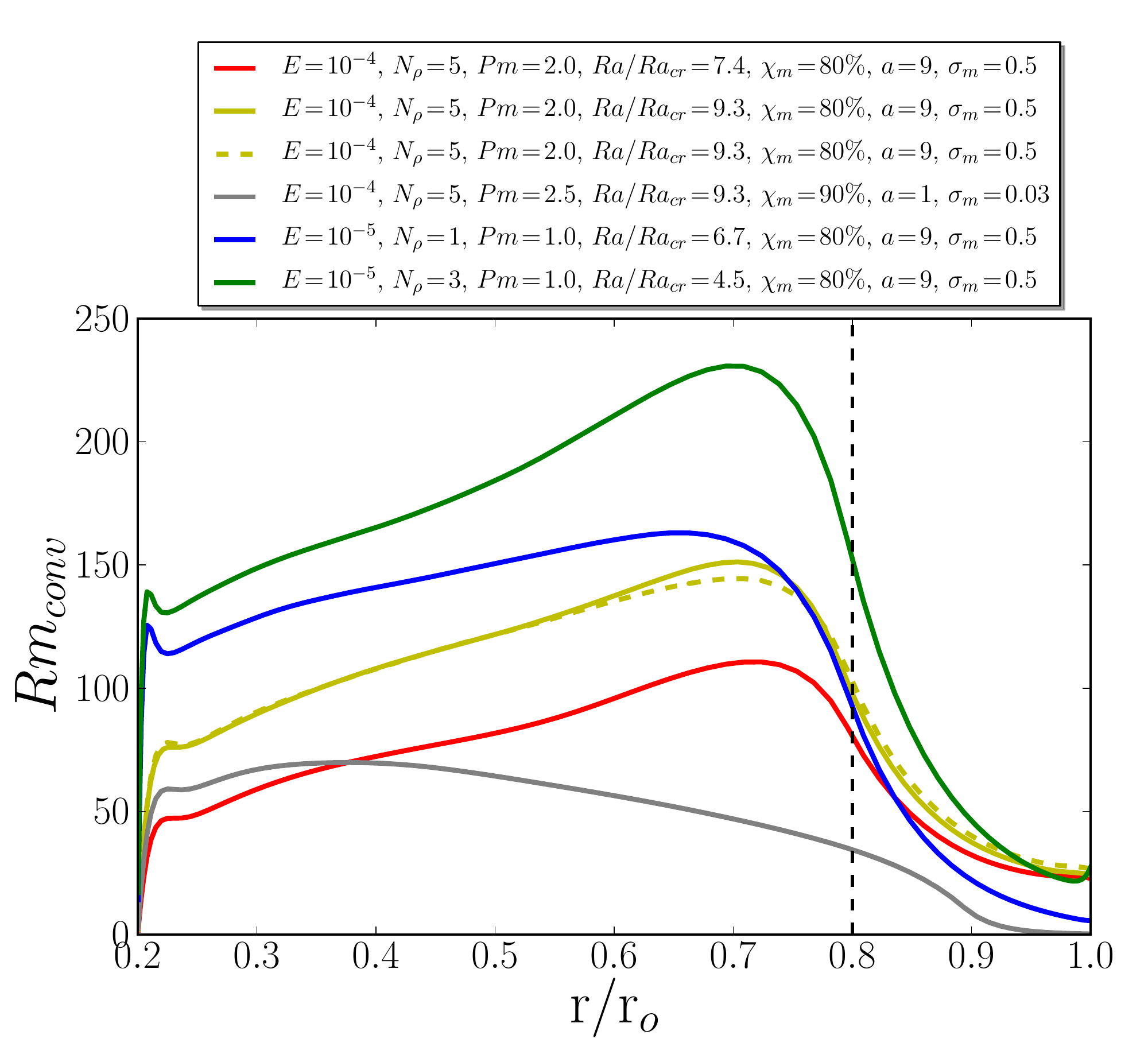}
      }
\caption{\small The radial profiles of convective magnetic Reynolds number averaged over time for
the cases displayed in Fig.~\ref{fig:zonalflow} (cases \tablecases{50} in grey,
\tablecases{51} in red, \tablecases{60} in blue and \tablecases{66} in green, from Tab.~\ref{Tab2}).
The two additional yellow cases
(\tablecases{53d/m} in Tab.~\ref{Tab2}) yield a higher $Ra$ and bistability,
at $N_\rho\!=\!5$ and $E=10^{-4}$.
\label{fig:Rms}}
\end{center}
\end{figure}

We also tested a more realistic profile that models the approximately linear
decrease of electrical conductivity in the metallic layer
(see Fig. \ref{fig:vconds}) and a steeper decrease at larger radii.
At $\chi_m\!=\!80$\%, $N_\rho\!=\!5$ and $Ra/Ra_c\!=\!9.3$ the solution is
bistable for our standard conductivity profile. For the more realistic profile we so far only found
a clearly dipole-dominated at the same supercriticality but we cannot exclude that the multipolar
case also exists.
\Cref{fig:Rms} compares the radial profiles of the convective magnetic Reynolds
number $Rm_{\scriptsize conv}$ for both profiles (yellow lines and grey line).
Being based on rms flows velocities that exclude
zonal winds, $Rm_{\scriptsize conv}$ is appropriate for characterizing
poloidal magnetic field
production. Numerical simulations suggest that a magnetic Reynolds number
larger than $50$ is required to support dynamo action \citep{Christensen06}.
For our standard electrical conductivity profiles, $Rm_{\scriptsize conv}$
typically falls below this value for radii beyond $r/r_o\!=\!0.85$ or $0.9$.
The linear decrease in the metallic layer, however, further reduces the
convective Reynolds number which is already very low at depth.
$Rm_{\scriptsize conv}$ values larger than $50$ are now restricted to the
inner region of $r/r_o\!<\!0.6$. A multipolar dynamo where the outer
parts of the shell play a sizeable role thus becomes unlikely.
In Jupiter, $Rm_{\scriptsize conv}$ is
generally significantly higher in the metallic region and only decreases
below the critical value for dynamo action in the molecular envelope.
We therefore refrained from further exploring
this profile since the decrease in magnetic Reynolds number artificially
limits the dynamo region.
 
\begin{figure*}[ht]
\begin{center}
{\centering
      \includegraphics[trim=0cm 0cm 1cm 1cm, height=1.9in]{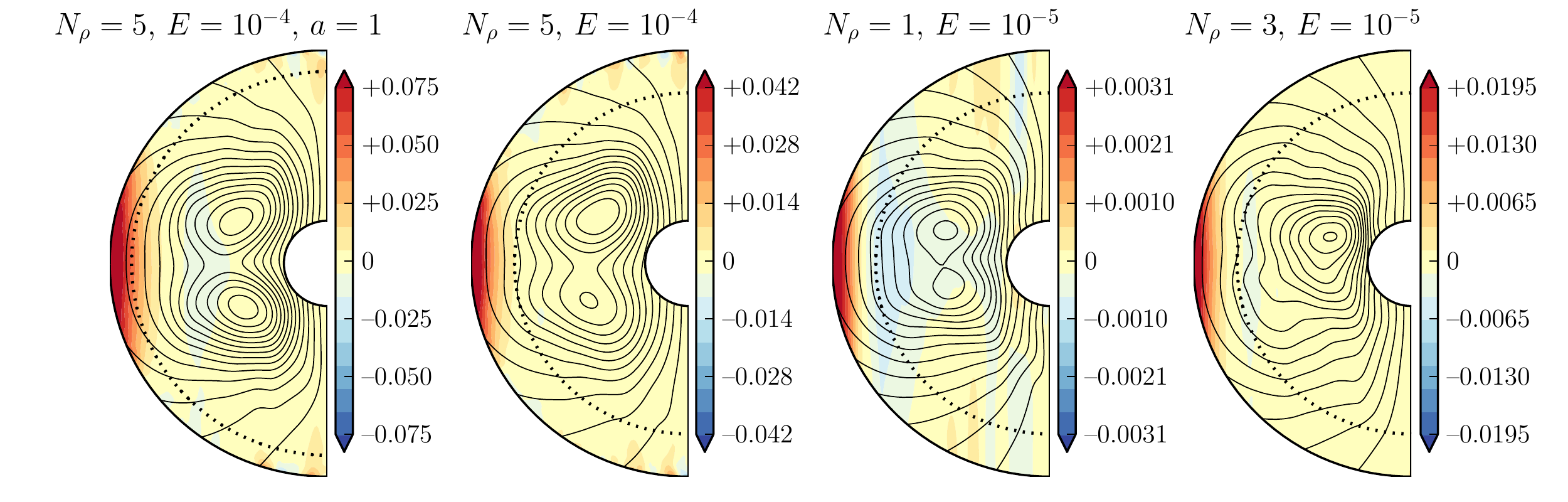}
    }
\caption{\small Azimuthal average of the zonal component of the flow.
The cases from Tab.~\ref{Tab2} displayed here are, from left to right: \tablecases{50} and
\tablecases{51} of $E\!=\!10^{-4}$, \tablecases{60} and \tablecases{66} of $E\!=\!10^{-5}$.
The poloidal field lines are plotted on top of the zonal velocity contours,
where solid lines are positive and dashed are negative values. The dotted lines correspond
to $\chi_m$. Zonal velocities are given in units of Rossby number calculated by
$Ro=u/(\Omega r_o)$.
\label{fig:vppolN1}}
\end{center}
\end{figure*}

\Cref{fig:vppolN1} shows zonal flows and axisymmetric poloidal field
lines for the dipole-dominated solutions at $E\!=\!10^{-4}$ with the modified electrical
conductivity profile (first panel from the left) and the standard profile (second panel).
The poloidal fields are very similar and produced at greater depth in
both cases. This explains why the low convective magnetic Reynolds number
in the outer part of the shell has little impact on the dynamo mechanism for
dipolar dominated solutions.
Once more, the weaker Lorentz force in the outer layer allows for
more vigorous zonal winds for the more realistic conductivity profile.

%LOWER EKMAN NUMBER:
How do the results at $E\!=\!10^{-5}$ fit into the picture we outlined above?
\Cref{fig:vppolN1} compares two $E\!=\!10^{-5}$ cases
at $N_{\rho}\!=\!1$ (third panel, case \tablecases{60}) and
$N_{\rho}\!=\!3$ (fourth panel, case \tablecases{66})
with the dipole-dominated solutions for the more realistic profile
(first panel, case \tablecases{50}) and for
our standard profile at $N_{\rho}\!=\!5$ and $E\!=\!10^{-4}$
(second panel, case \tablecases{51}, see also Fig.~\ref{fig:zonalflow}).
The magnetic field and zonal flow structures are very similar in all cases.
At both Ekman numbers, the relative amplitude of the retrograde jets
decreases with increasing $N_\rho$ (see also Tab.~\ref{Tab2}).
The absolute zonal flow amplitude, however, is significantly
smaller in all lower Ekman number models.
For example, the zonal flow Rossby number is
$Ro_{zon}\!=\!5.2\times10^{-3}$ in the $E\!=\!10^{-5}$ / $N_{\rho}\!=\!3$
case (\tablecases{66}) depicted in Fig.~\ref{fig:vppolN1}, but
$Ro_{zon}\!=\!1.4\times10^{-2}$ in the $E\!=\!10^{-4}$ / $N_{\rho}\!=\!5$
simulation (\tablecases{51}).%, one order of magnitude larger.

In non-magnetic free-slip simulations, the flow amplitude roughly scales with the
modified Rayleigh number $Ra^\star\!=\!Ra\,E^2/Pr$, as it has been shown
by \cite{Christensen02} for Boussinesq and \cite{Gastine12} for
anelastic models.
For example, \cite{Gastine12} suggest the dependence
$Ro\approx 0.165\,{Ra^\star}^{1.06}$.
This scaling describes an asymptotic behaviour for larger Rayleigh
numbers where zonal flows clearly dominate so that $Ro\approx Ro_{zon}$.
For the smaller Rayleigh numbers typically examined here, it may
only serve as a rough estimate for the zonal flow amplitude.
For case \tablecases{51} with $E\!=\!10^{-4}$, we have $Ra^\star\!=\!=0.4$ and the
scaling predicts $Ro_{zon}\!\approx\!6.2\times10^{-2}$.
For case \tablecases{66} with $E\!=\!10^{-5}$ and $Ra^\star\!=\!=0.04$, it predicts
$Ro_{zon}\!\approx\!5.4\times10^{-3}$.
Both values are not too far from the numerical results
$Ro_{zon}\!\approx\!1.4\times10^{-2}$ and $Ro_{zon}\!\approx\!5.2\times10^{-3}$,
respectively, which suggests that the difference in $Ra^\star$ is indeed
the main reason for the much weaker zonal flows at the lower Ekman number.

Because of the quadratic Ekman number dependence of $Ra^\star$,
$Ra$ has to be increased by two orders of magnitude to reach the
same zonal flow amplitudes in the $E\!=\!10^{-5}$
as in the $E\!=\!10^{-4}$ cases. This leads to larger $Ro_\ell$ values and
thus possibly multipolar fields \citep{Heimpel11}.
The Rayleigh number increase from $Ra/Ra_{cr}\!=\!10.0$
(case \tablecases{61}) to
$Ra/Ra_{cr}=16.7$ (case \tablecases{62}) at $N_\rho\!=\!1$ already
leads to a multipolar field while only doubling the zonal flow amplitude.

The similar Elsasser numbers in the dipole-dominated cases at both
Ekman numbers indicate that the Lorentz forces also have comparable
amplitudes. These forces have a much easier job to brake the systematically
weak zonal flows at $E\!=\!10^{-5}$, allowing a dipole-dominated field to
develop even at mild stratifications.
The more extensive parameter study at $E\!=\!10^{-4}$ suggests that
stronger stratifications should allow for more vigorous outer jets while
retaining dipole-dominated dynamo action.

\subsection{Dynamo Mechanism}
\label{dynamo}

GDW12 reported that the multipolar solutions with stronger zonal flows
are dynamos of an $\alpha\Omega$ or an $\alpha^2\Omega$ type.
Dynamos of the $\alpha^2$ type, on the other hand, are known to produce
dipole-dominated magnetic fields \citep{Olson99}.
The $\alpha$ stands for poloidal and toroidal field
production by local helical structures, while $\Omega$ stands for the
production of toroidal field by global zonal wind shear.
Following \cite{Brown11}, the $\Omega$-effect is given by
\begin{equation}
      \Omega=\overline{B}_r \frac{\partial}{\partial r}
      \bigg(\frac{\overline{u}_\phi}{r}\bigg)
      +\frac{\overline{B}_\theta\sin\theta}{r}\frac{\partial}
      {\partial\theta}\bigg(\frac{\overline{u}_\phi}{\sin\theta}\bigg)
\label{eq:omegaeffect}
\end{equation}
and describes the production of the axisymmetric
azimuthal magnetic field $\overline{B}_\phi$ which is purely toroidal.
%Overbars correspond to azimuthal averages here.
The mean ohmic diffusion of $\overline{B}_\phi$ is given by
\begin{equation}
      \mathrm{MD}=
        \tilde{\lambda}\nabla^2\overline{B}_\phi
        -\frac{\tilde{\lambda}\overline{B}_\phi}{r^2\sin^2\theta}+
        \bigg(\frac{\partial\tilde{\lambda}}{\partial r}\bigg)\bigg(\frac{1}{r}
        \frac{\partial r\overline{B}_\phi }{\partial r}\bigg)
      \textrm{.}
\label{eq:ohmicdiff}
\end{equation}

\begin{figure}[ht]
\begin{center}
{\centering
      \includegraphics[trim=0cm 0cm 0.5cm 0cm, height=5.4in]{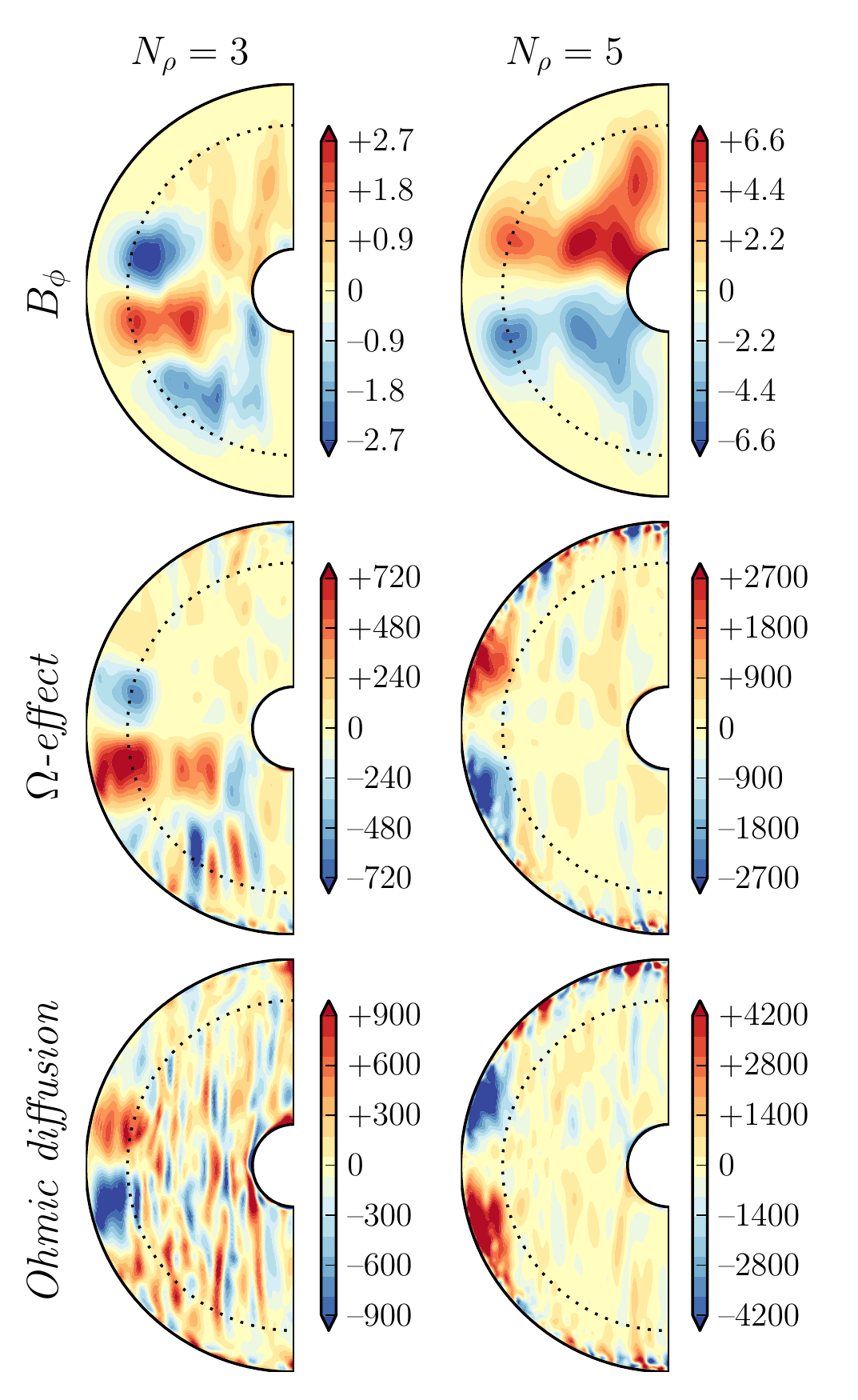}
}
\caption{\small From top to bottom, azimuthal averages of the toroidal
component of the
magnetic field, production of the toroidal field by $\Omega$-effect and ohmic diffusion.
The three left panels correspond to $N_\rho\!=\!3$, $Ra/Ra_{cr}\!=\!4.3$ and
the three right panels to $N_\rho\!=\!5$, $Ra/Ra_{cr}\!=\!7.4$
(cases \tablecases{33} and \tablecases{51} from Tab.~\ref{Tab2}), both cases
belong to the red $\chi_m=80\%$ line of Fig.~\ref{fig:DipvsNrho}.
\label{fig:omegas}}
\hspace{-20pt}
\end{center}
\end{figure}

\Cref{fig:omegas} compares $\overline{B}_\phi$, $\Omega$ and MD for
two $\chi_m\!=\!80$\% cases, a multipolar solution at $N_\rho\!=\!3$ (left)
and dipole dominated solution at $N_\rho\!=\!5$ (right).
Both cases were depicted previously in Figs.~\ref{fig:zonalflow}
and \ref{fig:EMagvsRad}.
For the multipolar solution, the $\Omega$-effect plays
an important role, as demonstrated by the high degree of correlation
with the azimuthal field over the shell.
For the dipole-dominated solution, however, the
$\Omega$-effect is only strong in the weakly conducting region where
it is effectively balanced by the large ohmic diffusion.
The change in field geometry is thus once more coupled to a switch
from an $\alpha\Omega$ or $\alpha^2\Omega$ mechanism at weaker stratifications
to an $\alpha^2$ mechanism at stronger density stratifications.

\section{Conclusions}
\label{conclusions}

We ran a suite of dynamo simulations with an electrical conductivity
profile geared to combine the dynamics of the metallic and the molecular
hydrogen layers of the gas giants in one integrated model.
In most models, the conductivity is assumed to remain constant over
the inner part of the shell representing the metallic hydrogen region.
Beyond a relative radius of $\chi_m$, it decays exponentially with radius,
to model the molecular envelope. The use of the anelastic MHD code
allowed us to also study the effects of density stratification. Free-slip outer
boundary and no-slip inner boundary conditions seem appropriate for the gas
giants and were used in the majority of our simulations.

In GDW12, we had explored the dynamo action for homogeneous electrical
conductivity in an otherwise identical setup. Dipole-dominated solutions
were only found for mild stratifications and local Rossby numbers
below $Ro_{\ell c}=0.08$. The respective solution branch is characterized
by weak zonal winds and coexists with a second branch with weaker multipolar
magnetic fields but stronger zonal winds at identical parameters.
This indicates a competition between zonal winds and dipolar magnetic fields
\citep{Simitev09,Schrinner12,Gastine12a}.
Translated to Jupiter and Saturn, these simulations would predict
multipolar magnetic fields, should the observed zonal winds reach into
the dynamo region. The strong stratification within the gas giants should
also promote multipolar solutions.

We largely recover these results when the outer weakly conducting
layer occupies only $5$ or $10$ percent in radius
($\chi_m\!=\!95\%$ and $\chi_m\!=\!90\%$). However, the critical
local Rossby number, below which dipole-dominated solutions are possible,
decreases to $Ro_{\ell c}\!\approx\!0.04$.
\cite{Gomez10} already showed that even very thin weakly conducting outer
layers promote multipolar magnetic field configurations.
They speculate that the separation of the Ekman and Hartmann boundary
layers may play a role in their models with no-slip boundaries, although
this explanation is difficult to apply for the free-slip
models predominantly explored here.
More research is required to clarify this point in the future,
specially concerning no-slip boundaries.

For a thicker weakly conducting outer layer covering the outer 20 or 30\% in
radius ($\chi_m\!=\!80$\% or $70$\%), the volume over which the Lorentz
forces can act to balance Reynolds stresses is more significantly reduced.
The competition between zonal wind and stronger dipolar fields thus
becomes even more of an issue.
At an Ekman number of $E\!=\!10^{-4}$, the mean
zonal winds tend to be relatively strong even at low Rayleigh numbers.
Dipole-dominated dynamo action is nevertheless possible in the deeper interior of
strongly stratified models, where the zonal flows remain relatively weak even
in the non-magnetic case. Alternatively, dipole-dominated solutions are found
for larger magnetic Prandtl numbers which help to keep zonal flows at bay
by increasing Lorentz forces. In the dipole-dominated
solutions, the zonal winds are then mainly restricted to a fierce
prograde jet that resides within the weakly conducting outer envelope.

At the lower Ekman number of $E\!=\!10^{-5}$, dipole-dominated magnetic fields
can even be maintained at weak stratifications because the zonal flow amplitudes
are lower than at $E\!=\!10^{-4}$.
The peak velocity of Jupiter's equatorial jet is around $Ro_{ej}\!=\!1.1\times10^{-2}$
\citep{Vasavada05} and  about $Ro_{ej}\!=\!5.0\times10^{-2}$ for Saturn \citep{Choi09}.
For example, the $E=10^{-4}$ case illustrated in Fig.~\ref{fig:vppolN1}
(second panel from the left) reaches $Ro_{ej}\!=\!5.3\times10^{-2}$ which
is somewhat too high for Jupiter.
The $E\!=\!10^{-5}$ simulation depicted in the
same figure has $Ro_{ej}\!=\!2.4\times10^{-2}$ at the lower stratification
of $N_{\rho}\!=\!3$.
The amplitude of the equatorial jet decreases with
Ekman number and increases with density stratification.
We speculate that the higher stratifications within the
gas giants may allow to reach appropriate zonal jet amplitudes at
the much lower realistic Ekman number, around $E_J\!\sim\!5\times10^{-19}$
\citep{French12}, while retaining dipole-dominated dynamo action.

The number of zonal jets is much smaller in our simulations than for the gas giants.
Also, the strong decrease in the zonal flow amplitude from the equatorial to
the flanking jets, that is necessary to retain dipole-dominated
dynamo action in our models, is not compatible with the observations for Jupiter.
A dipolar configuration nevertheless seems possible should the higher latitude
jets remain too shallow to interfere with the deeper dynamo process.
The equatorial jet does not pose a problem in this respect because
it can reside completely within the lower conductivity envelope.

An argument against deep reaching winds is that the associated strong
$\Omega$-effect and Ohmic dissipation may not be compatible with
Jupiter's observed luminosity \citep{Liu08}.
A first analysis of our results confirms that the $\Omega$-effect
and associated Ohmic dissipation can be significant.
\cite{Glatz08} argues that the magnetic field may assume a configuration
where the poloidal field lines are aligned with the rotation axis in
regions of strong zonal flow shear. Since the shear is perpendicular to
the rotation axis, this would minimize the $\Omega$-effect and related Ohmic
dissipation.
\Cref{fig:vppolN1} illustrates that the field lines indeed approach
such an alignment in the very outer part of the shell where the electrical
conductivity is still important. The Ohmic dissipation nevertheless remains
significant in all our simulations with strong zonal flows. Further
investigation is necessary to quantify this effect and extrapolate
it to the planetary situation.

Any problems related to ohmic dissipation and dipolar dynamo action would
not be an issue when stronger zonal winds remain confined to a thin outer
envelope with ($\chi_m\!\geq\!96\%$), where the electrical
conductivity remains small enough \citep{Liu08}.
In our simulations, however, all the stronger jets obey the
Taylor-Proudman theorem and reach through the planet.
Shallow jets have been found by \cite{Kaspi09},
who use a different anelastic approximation and a different
internal heating mode. Further investigations
are required to clarify which specific model features influence
the depth on the zonal jets.

\section*{Acknowledgements}

All the computations have been carried out in the GWDG computer facilities in
G\"ottingen, in the Norddeutscher Verbund für Hoch- und Höchstleistungsrechnen (HLRN)
in Hannover and in the Max-Planck-Institut für Sonnensystemforschung.
The authors would like to specially thank the reviewers and editor for the much helpful
comments and suggestions and also HLRN for providing the possibility of
carrying out demanding "last minute" simulations that allowed improvement of the work.
This work was supported by the Special Priority Program 1488
(PlanetMag, {http://www.planetmag.de}) of the German Science Foundation.

%\section*{References}
%\label{refenrences}

\bibliographystyle{elsarticle-harv}

{\small
%{\tiny

\onecolumn
\begin{center}
\begin{longtable}{ccccccccccccccc}

\caption{Summary of the time-averaged results\\
\small{
$^*$no-slip top boundary\\
$^{(1)}a\!=\!25$\\
$^{(2)}a\!=\!1,\,\sigma_m=0.03$\\
$^{**}$case from \cite{Heimpel11} with $\eta\!=\!0.35$
\label{Tab2}}}\\

\hline
Model & $\begin{matrix}\chi_m\\(\%)\end{matrix}$ & $N_{\rho}$ & $\dfrac{Ra}{Ra_{cr}}$ & $E$ & $Pm_i$ & $N_r\!\times\!\ell_{max}$ & $f_{dip}$ & $S\!D_{dip}$ & $Ro_{zon}$ & $Z$ & $Rm$ & $Ro_{\ell}$ & $\Lambda$ & $\tau$ \\ %& $\dfrac{E_{kin\,ax\,\phi}}{E_{kin}}$ &
\hline
\vspace{-9.5pt}
 & & & \\
\hline
\endfirsthead

\hline
Model & $\begin{matrix}\chi_m\\(\%)\end{matrix}$ & $N_{\rho}$ & $\dfrac{Ra}{Ra_{cr}}$ & $E$ & $Pm_i$ & $N_r\!\times\!\ell_{max}$ & $f_{dip}$ & $S\!D_{dip}$ & $Ro_{zon}$ & $Z$ & $Rm$ & $Ro_{\ell}$ & $\Lambda$ & $\tau$ \\
\hline
\vspace{-9.5pt}
 & & & \\
\hline
\endhead

\hline
\multicolumn{15}{r}{{Continued on next page}} \\
\hline
\endfoot
\endlastfoot

\vspace{-8pt}
 & & & \\
01d & 95 & 0.0 & 11.5 & $10^{-4}$ & $2.0$ & 73$\times$85 & $8.45\times 10^{-1}$ & $4.33\times 10^{-2}$ & $1.64\times 10^{-3}$ & 0.04 & 130 & $2.72\times 10^{-2}$ & $1.709$ & $3.3$ \\
01m & 95 & 0.0 & 11.5 & $10^{-4}$ & $2.0$ & 73$\times$106 & $1.12\times 10^{-1}$ & $8.10\times 10^{-2}$ & $5.59\times 10^{-3}$ & 0.36 & 157 & $2.66\times 10^{-2}$ & $0.575$ & $2.1$ \\
02 & 95 & 0.0 & 17.2 & $10^{-4}$ & $2.0$ & 73$\times$106 & $6.49\times 10^{-2}$ & $6.72\times 10^{-2}$ & $8.66\times 10^{-3}$ & 0.37 & 236 & $4.26\times 10^{-2}$ & $1.236$ & $1.8$ \\
03 & 95 & 0.0 & 23.0 & $10^{-4}$ & $2.0$ & 73$\times$106 & $2.81\times 10^{-2}$ & $2.98\times 10^{-2}$ & $1.02\times 10^{-2}$ & 0.35 & 290 & $5.43\times 10^{-2}$ & $1.936$ & $1.4$ \\
04$^*$ & 95 & 0.0 & 23.0 & $10^{-4}$ & $2.0$ & 73$\times$106 & $7.50\times 10^{-1}$ & $3.53\times 10^{-2}$ & $2.80\times 10^{-3}$ & 0.04 & 239 & $6.37\times 10^{-2}$ & $2.008$ & $1.6$ \\

\hline
05 & 90 & 0.0 & 9.2 & $10^{-4}$ & $5.0$ & 61$\times$64 & $6.21\times 10^{-1}$ & $1.81\times 10^{-1}$ & $1.62\times 10^{-3}$ & 0.07 & 242 & $1.78\times 10^{-2}$ & $3.892$ & $3.2$ \\
06 & 90 & 0.0 & 10.3 & $10^{-4}$ & $2.0$ & 73$\times$85 & $3.88\times 10^{-2}$ & $5.14\times 10^{-2}$ & $6.73\times 10^{-3}$ & 0.57 & 130 & $1.86\times 10^{-2}$ & $0.262$ & $1.5$ \\
07 & 90 & 0.0 & 11.5 & $10^{-4}$ & $2.0$ & 73$\times$106 & $4.79\times 10^{-2}$ & $5.07\times 10^{-2}$ & $6.93\times 10^{-3}$ & 0.50 & 153 & $2.31\times 10^{-2}$ & $0.432$ & $2.9$ \\
08$^*$ & 90 & 0.0 & 11.5 & $10^{-4}$ & $2.0$ & 73$\times$106 & $8.64\times 10^{-1}$ & $2.48\times 10^{-2}$ & $1.45\times 10^{-3}$ & 0.04 & 114 & $2.59\times 10^{-2}$ & $1.618$ & $2.9$ \\
09 & 90 & 0.0 & 23.0 & $10^{-4}$ & $2.0$ & 73$\times$106 & $7.06\times 10^{-2}$ & $7.62\times 10^{-2}$ & $1.12\times 10^{-2}$ & 0.38 & 270 & $5.21\times 10^{-2}$ & $1.857$ & $1.1$ \\
10$^*$ & 90 & 0.0 & 23.0 & $10^{-4}$ & $2.0$ & 73$\times$106 & $1.16\times 10^{-1}$ & $8.61\times 10^{-2}$ & $5.09\times 10^{-3}$ & 0.10 & 241 & $6.49\times 10^{-2}$ & $0.524$ & $2.2$ \\

\hline
11 & 80 & 0.0 & 11.5 & $10^{-4}$ & $2.0$ & 73$\times$106 & $5.30\times 10^{-2}$ & $5.81\times 10^{-2}$ & $8.98\times 10^{-3}$ & 0.65 & 140 & $1.91\times 10^{-2}$ & $0.255$ & $3.6$ \\
12 & 80 & 0.0 & 11.5 & $10^{-4}$ & $10.0$ & 73$\times$106 & $8.03\times 10^{-1}$ & $5.06\times 10^{-2}$ & $2.35\times 10^{-3}$ & 0.09 & 403 & $2.81\times 10^{-2}$ & $6.480$ & $1.6$ \\
13$^*$ & 80 & 0.0 & 11.5 & $10^{-4}$ & $10.0$ & 73$\times$106 & $7.60\times 10^{-1}$ & $5.92\times 10^{-2}$ & $1.45\times 10^{-3}$ & 0.03 & 395 & $2.52\times 10^{-2}$ & $6.728$ & $1.3$ \\
14 & 80 & 0.0 & 17.2 & $10^{-4}$ & $2.0$ & 73$\times$106 & $8.66\times 10^{-2}$ & $1.09\times 10^{-1}$ & $1.38\times 10^{-2}$ & 0.64 & 206 & $3.06\times 10^{-2}$ & $0.727$ & $1.4$ \\
15$^*$ & 80 & 0.0 & 20.7 & $10^{-4}$ & $2.0$ & 73$\times$85 & $6.26\times 10^{-2}$ & $9.18\times 10^{-2}$ & $5.05\times 10^{-3}$ & 0.11 & 162 & $5.66\times 10^{-2}$ & $0.195$ & $1.1$ \\
16 & 80 & 0.0 & 23.0 & $10^{-4}$ & $2.0$ & 73$\times$106 & $4.78\times 10^{-2}$ & $5.62\times 10^{-2}$ & $1.86\times 10^{-2}$ & 0.62 & 276 & $4.25\times 10^{-2}$ & $1.178$ & $3.4$ \\
17$^*$ & 80 & 0.0 & 23.0 & $10^{-4}$ & $2.0$ & 73$\times$106 & $3.40\times 10^{-2}$ & $4.50\times 10^{-2}$ & $5.00\times 10^{-3}$ & 0.09 & 177 & $5.82\times 10^{-2}$ & $0.232$ & $1.5$ \\
18 & 80 & 0.0 & 45.9 & $10^{-4}$ & $2.0$ & 81$\times$170 & $1.11\times 10^{-1}$ & $3.64\times 10^{-2}$ & $3.12\times 10^{-2}$ & 0.62 & 454 & $7.55\times 10^{-2}$ & $2.401$ & $1.0$ \\

\hline
19d & 95 & 1.0 & 4.1 & $10^{-4}$ & $2.0$ & 73$\times$85 & $8.29\times 10^{-2}$ & $4.76\times 10^{-2}$ & $1.27\times 10^{-3}$ & 0.05 & 90 & $2.61\times 10^{-2}$ & $1.227$ & $1.2$ \\
19m & 95 & 1.0 & 4.1 & $10^{-4}$ & $2.0$ & 73$\times$85 & $9.74\times 10^{-2}$ & $8.93\times 10^{-2}$ & $2.86\times 10^{-3}$ & 0.21 & 103 & $2.63\times 10^{-2}$ & $0.399$ & $1.2$ \\
20d & 95 & 1.0 & 5.2 & $10^{-4}$ & $2.0$ & 73$\times$106 & $8.36\times 10^{-1}$ & $3.49\times 10^{-2}$ & $1.60\times 10^{-3}$ & 0.04 & 121 & $3.77\times 10^{-2}$ & $2.210$ & $2.6$ \\
20m & 95 & 1.0 & 5.2 & $10^{-4}$ & $2.0$ & 73$\times$106 & $5.24\times 10^{-2}$ & $6.01\times 10^{-2}$ & $3.82\times 10^{-3}$ & 0.18 & 145 & $4.00\times 10^{-2}$ & $0.754$ & $2.4$ \\
21 & 95 & 1.0 & 7.8 & $10^{-4}$ & $2.0$ & 73$\times$85 & $3.86\times 10^{-2}$ & $4.81\times 10^{-2}$ & $6.83\times 10^{-3}$ & 0.21 & 234 & $6.26\times 10^{-2}$ & $1.849$ & $1.1$ \\

\hline
22d & 90 & 1.0 & 5.2 & $10^{-4}$ & $2.0$ & 73$\times$85 & $8.74\times 10^{-1}$ & $2.83\times 10^{-2}$ & $1.88\times 10^{-3}$ & 0.06 & 103 & $3.31\times 10^{-2}$ & $1.881$ & $1.8$ \\
22m & 90 & 1.0 & 5.2 & $10^{-4}$ & $2.0$ & 73$\times$85 & $8.12\times 10^{-2}$ & $1.08\times 10^{-1}$ & $4.63\times 10^{-3}$ & 0.26 & 126 & $3.47\times 10^{-2}$ & $0.602$ & $1.4$ \\
23d$^{*}$ & 90 & 1.0 & 5.2 & $10^{-4}$ & $2.0$ & 73$\times$85 & $8.36\times 10^{-1}$ & $3.68\times 10^{-2}$ & $1.66\times 10^{-3}$ & 0.04 & 110 & $3.64\times 10^{-2}$ & $2.103$ & $2.0$ \\
23m$^{*}$ & 90 & 1.0 & 5.2 & $10^{-4}$ & $2.0$ & 73$\times$85 & $1.01\times 10^{-1}$ & $9.06\times 10^{-2}$ & $2.01\times 10^{-3}$ & 0.05 & 131 & $4.41\times 10^{-2}$ & $0.358$ & $4.4$ \\
24 & 90 & 1.0 & 7.8 & $10^{-4}$ & $2.0$ & 73$\times$106 & $1.13\times 10^{-1}$ & $1.29\times 10^{-1}$ & $8.55\times 10^{-3}$ & 0.30 & 204 & $5.46\times 10^{-2}$ & $1.491$ & $1.8$ \\

\hline
25 & 80 & 1.0 & 5.2 & $10^{-4}$ & $2.0$ & 73$\times$106 & $3.53\times 10^{-2}$ & $4.23\times 10^{-2}$ & $8.08\times 10^{-3}$ & 0.58 & 95 & $2.25\times 10^{-2}$ & $0.252$ & $4.3$ \\
26 & 80 & 1.0 & 10.3 & $10^{-4}$ & $2.0$ & 73$\times$106 & $8.79\times 10^{-2}$ & $8.43\times 10^{-2}$ & $2.19\times 10^{-2}$ & 0.65 & 224 & $5.19\times 10^{-2}$ & $1.172$ & $2.0$ \\

\hline
27 & 95 & 2.0 & 2.9 & $10^{-4}$ & $2.0$ & 73$\times$106 & $4.20\times 10^{-2}$ & $3.77\times 10^{-2}$ & $4.34\times 10^{-3}$ & 0.28 & 107 & $3.66\times 10^{-2}$ & $0.503$ & $2.4$ \\

\hline
28 & 95 & 3.0 & 3.2 & $10^{-4}$ & $2.0$ & 73$\times$106 & $2.10\times 10^{-2}$ & $2.04\times 10^{-2}$ & $7.02\times 10^{-3}$ & 0.29 & 158 & $5.68\times 10^{-2}$ & $0.933$ & $3.8$ \\

\hline
29 & 90 & 3.0 & 3.2 & $10^{-4}$& $2.0$ & 73$\times$85 & $5.64\times 10^{-2}$ & $6.69\times 10^{-2}$ & $9.05\times 10^{-3}$ & 0.45 & 124 & $4.16\times 10^{-2}$ & $0.563$ & $2.4$ \\

\hline
30 & 80 & 3.0 & 3.2 & $10^{-4}$ & $2.0$ & 73$\times$85 & $1.19\times 10^{-1}$ & $1.03\times 10^{-1}$ & $1.38\times 10^{-2}$ & 0.72 & 76 & $2.38\times 10^{-2}$ & $0.197$ & $3.7$ \\
31 & 80 & 3.0 & 3.2 & $10^{-4}$ & $6.0$ & 121$\times$106 & $2.33\times 10^{-1}$ & $1.98\times 10^{-1}$ & $1.04\times 10^{-2}$ & 0.52 & 215 & $3.01\times 10^{-2}$ & $1.420$ & $1.0$ \\
32$^*$ & 80 & 3.0 & 3.2 & $10^{-4}$ & $2.0$ & 73$\times$106 & $1.08\times 10^{-1}$ & $9.02\times 10^{-2}$ & $6.75\times 10^{-3}$ & 0.23 & 90 & $3.71\times 10^{-2}$ & $0.198$ & $1.5$ \\
33 & 80 & 3.0 & 4.3 & $10^{-4}$ & $2.0$ & 73$\times$106 & $2.80\times 10^{-1}$ & $1.80\times 10^{-1}$ & $2.06\times 10^{-2}$ & 0.67 & 134 & $4.25\times 10^{-2}$ & $0.529$ & $3.9$ \\
34$^{(1)}$ & 80 & 3.0 & 4.3 & $10^{-4}$ & $2.0$ & 121$\times$106 & $2.34\times 10^{-1}$ & $1.37\times 10^{-1}$ & $2.18\times 10^{-2}$ & 0.71 & 128 & $3.90\times 10^{-2}$ & $0.453$ & $2.6$ \\
35 & 80 & 3.0 & 4.3 & $10^{-4}$ & $6.0$ & 145$\times$106 & $2.27\times 10^{-2}$ & $2.52\times 10^{-2}$ & $1.45\times 10^{-2}$ & 0.45 & 341 & $4.88\times 10^{-2}$ & $2.446$ & $0.7$ \\
36$^*$ & 80 & 3.0 & 4.3 & $10^{-4}$ & $2.0$ & 73$\times$106 & $7.52\times 10^{-2}$ & $7.68\times 10^{-2}$ & $8.55\times 10^{-3}$ & 0.19 & 137 & $5.62\times 10^{-2}$ & $0.629$ & $1.6$ \\
37 & 80 & 3.0 & 8.6 & $10^{-4}$ & $2.0$ & 81$\times$170 & $5.25\times 10^{-2}$ & $3.55\times 10^{-2}$ & $3.47\times 10^{-2}$ & 0.55 & 258 & $8.86\times 10^{-2}$ & $2.164$ & $1.2$ \\

\hline
38 & 70 & 3.0 & 4.3 & $10^{-4}$ & $2.0$ & 73$\times$106 & $2.03\times 10^{-1}$ & $2.01\times 10^{-1}$ & $2.52\times 10^{-2}$ & 0.77 & 90 & $2.22\times 10^{-2}$ & $0.224$ & $1.6$ \\
39 & 70 & 3.0 & 4.3 & $10^{-4}$ & $6.0$ & 129$\times$106 & $3.07\times 10^{-1}$ & $2.19\times 10^{-1}$ & $2.25\times 10^{-2}$ & 0.71 & 241 & $2.52\times 10^{-2}$ & $1.035$ & $0.8$ \\

\hline
40 & 95 & 4.0 & 5.5 & $10^{-4}$ & $2.0$ & 81$\times$170 & $4.07\times 10^{-3}$ & $4.21\times 10^{-3}$ & $9.46\times 10^{-3}$ & 0.27 & 168 & $8.57\times 10^{-2}$ & $1.185$ & $1.3$ \\

\hline
41 & 80 & 4.0 & 4.4 & $10^{-4}$ & $2.0$ & 81$\times$170 & $1.07\times 10^{-2}$ & $2.08\times 10^{-2}$ & $1.52\times 10^{-2}$ & 0.71 & 69 & $1.78\times 10^{-2}$ & $0.072$ & $3.0$ \\
42 & 80 & 4.0 & 5.5 & $10^{-4}$ & $2.0$ & 81$\times$170 & $4.40\times 10^{-1}$ & $2.76\times 10^{-1}$ & $2.00\times 10^{-2}$ & 0.66 & 102 & $3.27\times 10^{-2}$ & $0.297$ & $2.3$ \\
43 & 80 & 4.0 & 5.5 & $10^{-4}$ & $4.0$ & 97$\times$170 & $8.06\times 10^{-1}$ & $1.77\times 10^{-2}$ & $1.33\times 10^{-2}$ & 0.43 & 156 & $3.90\times 10^{-2}$ & $1.672$ & $1.1$ \\
44 & 80 & 4.0 & 8.8 & $10^{-4}$ & $2.0$ & 81$\times$170 & $2.43\times 10^{-1}$ & $2.26\times 10^{-1}$ & $3.01\times 10^{-2}$ & 0.61 & 184 & $5.83\times 10^{-2}$ & $0.912$ & $1.2$ \\

\hline
45 & 70 & 4.0 & 6.6 & $10^{-4}$ & $2.0$ & 81$\times$170 & $3.38\times 10^{-1}$ & $2.65\times 10^{-1}$ & $3.24\times 10^{-2}$ & 0.78 & 93 & $2.23\times 10^{-2}$ & $0.172$ & $1.5$ \\
46 & 70 & 4.0 & 6.6 & $10^{-4}$ & $4.0$ & 81$\times$170 & $9.32\times 10^{-1}$ & $1.09\times 10^{-2}$ & $2.28\times 10^{-2}$ & 0.61 & 118 & $2.75\times 10^{-2}$ & $1.717$ & $1.1$ \\
47 & 70 & 4.0 & 8.8 & $10^{-4}$ & $2.0$ & 81$\times$170 & $4.16\times 10^{-1}$ & $2.40\times 10^{-1}$ & $4.37\times 10^{-2}$ & 0.79 & 129 & $3.01\times 10^{-2}$ & $0.331$ & $1.7$ \\

\hline
48 & 90 & 5.0 & 7.4 & $10^{-4}$ & $2.0$ & 97$\times$170 & $1.18\times 10^{-3}$ & $1.83\times 10^{-3}$ & $1.16\times 10^{-2}$ & 0.31 & 140 & $5.61\times 10^{-2}$ & $0.735$ & $1.1$ \\
49 & 90 & 5.0 & 9.3 & $10^{-4}$ & $2.0$ & 97$\times$170 & $4.85\times 10^{-3}$ & $2.27\times 10^{-3}$ & $1.32\times 10^{-2}$ & 0.28 & 181 & $7.06\times 10^{-2}$ & $1.069$ & $0.7$ \\
50$^{(2)}$ & 90 & 5.0 & 9.3 & $10^{-4}$ & $2.5$ & 81$\times$192 & $9.63\times 10^{-1}$ & $8.32\times 10^{-3}$ & $2.38\times 10^{-2}$ & 0.57 & 50 & $5.82\times 10^{-2}$ & $0.468$ & $1.1$ \\

\hline
51 & 80 & 5.0 & 7.4 & $10^{-4}$ & $2.0$ & 81$\times$170 & $9.02\times 10^{-1}$ & $1.40\times 10^{-2}$ & $1.37\times 10^{-2}$ & 0.38 & 69 & $3.33\times 10^{-2}$ & $0.643$ & $2.9$ \\
52d$^{(1)}$ & 80 & 5.0 & 7.4 & $10^{-4}$ & $2.0$ & 97$\times$170 & $9.43\times 10^{-1}$ & $8.77\times 10^{-3}$ & $2.02\times 10^{-2}$ & 0.60 & 58 & $2.82\times 10^{-2}$ & $0.319$ & $1.0$ \\
52m$^{(1)}$ & 80 & 5.0 & 7.4 & $10^{-4}$ & $2.0$ & 97$\times$170 & $4.73\times 10^{-1}$ & $2.35\times 10^{-1}$ & $2.59\times 10^{-2}$ & 0.72 & 100 & $2.09\times 10^{-2}$ & $0.245$ & $1.6$ \\
53d & 80 & 5.0 & 9.3 & $10^{-4}$ & $2.0$ & 81$\times$170 & $9.03\times 10^{-1}$ & $8.03\times 10^{-3}$ & $1.69\times 10^{-2}$ & 0.37 & 94 & $4.62\times 10^{-2}$ & $1.074$ & $1.3$ \\
53m & 80 & 5.0 & 9.3 & $10^{-4}$ & $2.0$ & 97$\times$170 & $5.43\times 10^{-1}$ & $2.25\times 10^{-1}$ & $2.36\times 10^{-2}$ & 0.56 & 111 & $3.94\times 10^{-2}$ & $0.470$ & $1.6$ \\
54 & 80 & 5.0 & 11.2 & $10^{-4}$ & $2.0$ & 97$\times$170 & $4.84\times 10^{-1}$ & $2.74\times 10^{-1}$ & $2.44\times 10^{-2}$ & 0.49 & 128 & $5.04\times 10^{-2}$ & $0.732$ & $0.7$ \\

\hline
55 & 70 & 5.0 & 11.2 & $10^{-4}$ & $2.0$ & 97$\times$170 & $4.07\times 10^{-1}$ & $3.06\times 10^{-1}$ & $3.86\times 10^{-2}$ & 0.72 & 95 & $2.41\times 10^{-2}$ & $0.250$ & $0.8$ \\

\hline
56 & 80 & 5.5 & 9.7 & $10^{-4}$ & $2.0$ & 97$\times$170 & $9.14\times 10^{-1}$ & $1.53\times 10^{-2}$ & $1.39\times 10^{-2}$ & 0.30 & 76 & $3.44\times 10^{-2}$ & $0.820$ & $1.4$ \\

\hline
\hline

57 & 95 & 0.0 & 10.0 & $10^{-5}$ & $1.0$ & 81$\times$133 & $8.61\times 10^{-1}$ & $1.64\times 10^{-2}$ & $2.91\times 10^{-4}$ & 0.06 & 98 & $7.05\times 10^{-3}$ & $0.214$ & $1.1$ \\

\hline
58 & 80 & 0.0 & 12.5 & $10^{-5}$ & $1.0$ & 81$\times$170 & $8.71\times 10^{-1}$ & $2.45\times 10^{-2}$ & $4.91\times 10^{-4}$ & 0.11 & 90 & $7.44\times 10^{-3}$ & $0.324$ & $1.0$ \\
59 & 80 & 0.0 & 20.8 & $10^{-5}$ & $1.0$ & 81$\times$170 & $8.52\times 10^{-1}$ & $2.69\times 10^{-2}$ & $1.07\times 10^{-3}$ & 0.17 & 138 & $1.37\times 10^{-2}$ & $0.969$ & $0.9$ \\

\hline
60 & 80 & 1.0 & 6.7 & $10^{-5}$ & $1.0$ & 81$\times$170 & $9.25\times 10^{-1}$ & $8.83\times 10^{-3}$ & $1.55\times 10^{-3}$ & 0.40 & 101 & $1.28\times 10^{-2}$ & $0.538$ & $1.1$ \\
61 & 80 & 1.0 & 10.0 & $10^{-5}$ & $1.0$ & 81$\times$170 & $8.87\times 10^{-1}$ & $1.83\times 10^{-2}$ & $3.06\times 10^{-3}$ & 0.41 & 158 & $1.97\times 10^{-2}$ & $1.301$ & $1.1$ \\
62 & 80 & 1.0 & 16.7 & $10^{-5}$ & $1.0$ & 97$\times$170 & $2.30\times 10^{-1}$ & $2.00\times 10^{-1}$ & $6.92\times 10^{-3}$ & 0.60 & 329 & $3.71\times 10^{-2}$ & $0.821$ & $0.4$ \\

\hline
63 & 80 & 2.0 & 3.6 & $10^{-5}$ & $1.0$ & 81$\times$170 & $8.64\times 10^{-1}$ & $1.34\times 10^{-2}$ & $1.92\times 10^{-3}$ & 0.54 & 84 & $1.31\times 10^{-2}$ & $0.234$ & $0.8$ \\
64 & 80 & 2.0 & 5.4 & $10^{-5}$ & $1.0$ & 81$\times$170 & $9.04\times 10^{-1}$ & $2.68\times 10^{-2}$ & $4.90\times 10^{-3}$ & 0.73 & 151 & $1.85\times 10^{-2}$ & $0.531$ & $1.5$ \\

\hline
65 & 80 & 3.0 & 3.4 & $10^{-5}$ & $1.0$ & 81$\times$170 & $9.35\times 10^{-1}$ & $9.74\times 10^{-3}$ & $4.94\times 10^{-3}$ & 0.81 & 102 & $1.41\times 10^{-2}$ & $0.276$ & $1.5$ \\
66 & 80 & 3.0 & 4.5 & $10^{-5}$ & $1.0$ & 81$\times$170 & $8.86\times 10^{-1}$ & $2.66\times 10^{-2}$ & $5.19\times 10^{-3}$ & 0.61 & 141 & $2.44\times 10^{-2}$ & $0.675$ & $1.0$ \\

\hline
\hline
\vspace{-10pt}
 & & & \\
67$^{**}$ & 80 & 0.0 & 7.2 & $10^{-5}$ & $3.0$ & 121$\times$170 & $8.40\times 10^{-1}$ & $9.11\times 10^{-3}$ & $1.82\times 10^{-3}$ & 0.50 & 306 & $1.20\times 10^{-2}$ & $2.471$ & $0.3$ \\

\hline

\end{longtable}
\end{center}
\twocolumn

}

%\end{linenumbers}

\end{document}